\documentclass[useAMS,usenatbib]{mn2e}

\usepackage{amssymb}
\usepackage{amsmath}
\usepackage{booktabs}
\PassOptionsToPackage{hyphens}{url}\usepackage{hyperref}
\usepackage{multirow}
\usepackage{times}
\usepackage[hyphenbreaks]{breakurl}
\usepackage[pdftex]{graphicx}

\title[X-ray variability of NGC\,3227]
  {X-ray variability analysis of a large series of {\it XMM-Newton} $+$ {\it NuSTAR} observations of NGC\,3227}
\author[A.P. Lobban]
  {A.P.~Lobban,$^{1,2}$\thanks{e-mail: \href{mailto:a.p.lobban1@keele.ac.uk}{alobban@sciops.esa.int}}
  T.J.~Turner$^3$, J.N.~Reeves$^3$, V.~Braito$^{4}$, L.~Miller$^{5}$
  \\
  $^1$Astrophysics Group, School of Physical and Geographical Sciences, Keele University, Keele, Staffordshire, ST5 5BG, U.K. \\
  $^2$European Space Agency (ESA), European Space Astronomy Centre (ESAC), E-28691 Villanueva de la Ca\~{n}ada, Madrid, Spain \\
  $^3$Department of Physics, University of Maryland Baltimore County, Baltimore, MD 21250 U.S.A \\
  $^4$INAF - Osservatorio Astronomico di Brera, Via Bianchi 46 I-23807 Merate (LC), Italy \\
  $^5$Department of Physics, University of Oxford, Keble Road, Oxford, OX1 3RH, U.K. \\
  }
\date{Accepted for publication in MNRAS on 6 April 2020}

\pagerange{\pageref{firstpage}--\pageref{lastpage}} \pubyear{2020}

\def\LaTeX{L\kern-.36em\raise.3ex\hbox{a}\kern-.15em
    T\kern-.1667em\lower.7ex\hbox{E}\kern-.125emX}

\begin{document}

\label{firstpage}

\maketitle

\begin{abstract}

We present a series of X-ray variability results from a long {\it XMM-Newton} $+$ {\it NuSTAR} campaign on the bright, variable AGN NGC\,3227.  We present an analysis of the light curves, showing that the source displays typically softer-when-brighter behaviour, although also undergoes significant spectral hardening during one observation which we interpret as due to an occultation event by a cloud of absorbing gas.  We spectrally decompose the data and show that the bulk of the variability is continuum-driven and, through rms variability analysis, strongly enhanced in the soft band.  We show that the source largely conforms to linear rms-flux behaviour and we compute X-ray power spectra, detecting moderate evidence for a bend in the power spectrum, consistent with existing scaling relations.  Additionally, we compute X-ray Fourier time lags using both the {\it XMM-Newton} and - through maximum-likelihood methods - {\it NuSTAR} data, revealing a strong low-frequency hard lag and evidence for a soft lag at higher frequencies, which we discuss in terms of reverberation models.

\end{abstract}


\section{Introduction} \label{sec:introduction}

The multiwavelength emission observed from active galactic nuclei (AGN) is thought to be powered by accretion onto a central supermassive black hole (SMBH).  These systems are observed to emit across the whole electromagnetic spectrum with components of both thermal and non-thermal emission contributing to the broad-band spectral energy distribution \citep{Shang11}.  The energy output of Seyfert galaxies is routinely seen to peak at ultraviolet (UV) wavelengths.  The dominant mechanism for this is widely considered to consist of thermal emission originating in material in the inner regions of a geometrically-thin, optically-thick accretion disc around the SMBH \citep{ShakuraSunyaev73}.  Depending on the accretion-flow properties, the region where the UV-emitting matter is located typically lies 10-1\,000\,$r_{\rm g}$\footnote{The definition of the gravitational radius is: $r_{\rm g} = GM_{\rm BH}/c^{2}$.} from the central SMBH.  Then, these thermally-emitted UV photons are likely responsible for generating X-ray emission via inverse-Compton scattering off hot electrons ($T \sim 10^{9}$\,K) in an optically-thin corona, most likely located within a few tens of $r_{\rm g}$ from the SMBH \citep{HaardtMaraschi93}.  This produces a continuum of X-ray emission, which commonly takes the form of a power law.

Our current generation of telescopes is unable to directly resolve the central regions of AGN.  This is due to their very compact nature coupled with their large distances from Earth.  Nevertheless, there are methods we can employ which allow us to indirectly infer information about the dominant structure, geometry and physical processes within these systems.  Variability studies, in particular, are powerful in this regard, allowing us to probe the observed X-ray variability in a number of model-independent ways --- e.g. analysis of the covariance and rms spectra.  Additionally, the frequency-dependence of the variability can be measured through such methods as estimating the power spectral density (PSD) and energy-dependent time delays (see below).  Curiously, the behavioural properties are often observed to be similar across a wide range of sources with their measured frequencies and amplitudes scaling roughly inversely with the SMBH mass (e.g. \citealt{Lawrence87, UttleyMcHardyPapadakis02, Markowitz03, Vaughan03, Papadakis04, McHardy04, McHardy06, Arevalo08}).  Additionally, the strength of the rms variability in the X-ray band is routinely observed to scale in an approximately linear manner with the flux of the source; i.e. the `rms-flux relation'.  This constitutes another common aspect of X-ray variability that appears to exist over a large range of time-scales and masses, observed in both AGN and X-ray binary (XRB) systems (e.g. \citealt{UttleyMcHardy01, Vaughan03, Gaskell04, UttleyMcHardyVaughan05}).

Time delays between the observed variations at different X-ray energies are commonly detected in bright, variable AGN (e.g. \citealt{PapadakisNandraKazanas01, VaughanFabianNandra03, McHardy07, LobbanAlstonVaughan14, Kara16, JinDoneWard17, Lobban18a, Lobban18b}).  In particular, hard X-ray variations are commonly seen to lag behind soft X-ray variations.  These are known as `hard lags' and their first detection was made in XRBs (e.g. Cygnus X-1: \citealt{Cui97, Nowak99}).  These typically occur at low variability frequencies ($\sim$10$^{-5} - 10^{-4}$\,Hz in most AGN), with the magnitude of the lag --- typically on the order of tens to hundreds of seconds in AGN, depending on the black hole mass --- often observed to be larger when the separation between the two energy bands is greater.  Given the similarities in their behaviour, it has been argued that the low-frequency hard lags that we observe in AGN and XRBs are analogous (e.g. \citealt{McHardy06}).  In this care, their properties --- e.g. time-scale, magnitude, etc --- are appropriately scaled according to the size of the emitting region 

In an attempt to explain the observed time delays, various models have been proposed.  These range from inverse Compton scattering of photons by the X-ray-producing corona (see \citealt{MiyamotoKitamoto89}) to reflection by the surface of the accretion disc \citep{KotovChurazovGilfanov01}.  A popular model was proposed by \citet{Lyubarskii97} and is known as the `propagating fluctuations' model.  Here, variations in the local mass accretion rate are responsible as they inwardly propagate through the disc.  Here, they then excite an extended corona of hot, relativistic electrons, ultimately producing the observed X-rays.  This picture works quite clearly for XRBs where X-rays are produced by the disc and an average hard delay is produced whereby stratification of the corona results in the inwardly-propagating fluctuations firstly exciting the outer regions of the corona, producing softer X-rays, before exciting the inner regions, driving emission with a harder spectrum.  However, a significant degree of complexity is added to the lag behaviour in AGN at higher frequencies.  Here, the time delays are often found to be reversed with more rapid variations in the soft X-ray band lagging behind the correlated variations at higher energies (i.e. `soft lags'; see \citealt{DeMarco13}).  A mechanism has been proposed by \citet{Miller10a} in which these lags emerge via scattering of the primary X-rays in more distant circumnuclear material.   In the context of this interpretation, the delayed signal is due to reverberation from absorbing material tens to hundreds of $r_{\rm g}$ from the central source (also see \citealt{Turner17, Mizumoto18, Mizumoto19}).  Here, the spectral shape of the delayed component is expected to be harder than the primary X-ray continuum with the `reflected' contribution increasing towards higher energies.  The observed time delay due to reverberation is diluted due to the presence of direct emission in the time series.  However, at higher energies, the relative contribution of delayed-to-direct emission increases, predicting a strong reverberation signal in the hard X-ray band; i.e. peaking in the {\it NuSTAR} bandpass at energies $> 10$\,keV (e.g. \citealt{Turner17}).  Alternatively, the soft lags are often discussed in terms of the reverberation signal from reflection of the primary X-rays off material very close to the central black hole (e.g. \citealt{ZoghbiUttleyFabian11, Fabian13}).  Also see \citet{Uttley14} for a review.

Here, we report on a large series of {\it XMM-Newton} $+$ {\it NuSTAR} observations of NGC\,3227, a nearby Seyfert 1.5 galaxy at a redshift of $z = 0.003859$ \citep{deVaucouleurs91} and an estimated luminosity distance of 20.3\,Mpc \citep{Mould00}.  The source is very bright in the X-ray band with a typical observed flux of $F_{\rm 0.3-10} \sim 6 \times 10^{-11}$\,erg\,cm$^{-2}$\,s$^{-1}$ from 0.3--10\,keV and has an estimated black hole mass of $M_{\rm BH} = 5.96^{+1.23}_{-1.36} \times 10^{6}$\,$M_{\odot}$ \citep{BentzKatz15}.  The source has been known to exhibit both bright, unabsorbed and low, absorbed states based on previous X-ray analyses (e.g. \citealt{LamerUttleyMcHardy03, Markowitz09, Markowitz14, RiversMarkowitzRothschild11}) with an outflowing warm absorber \citep{Beuchert15}.  Interestingly, the source is occasionally observed to undergo variable absorption events apparently due to clouds of gas occulting across the line of sight (e.g. \citealt{LamerUttleyMcHardy03, Beuchert15}).

This is the second in a series of papers on this co-ordinated {\it XMM-Newton} $+$ {\it NuSTAR} campaign.  Previously, in \citet{Turner18}, we reported on a rapid occultation event occurring towards the end of the observing campaign.  We detected significant spectral hardening of the source coupled with a measurable increase in the depth of the unresolved transition array (UTA), thanks to the high-resolution Reflection Grating Spectrometer (RGS) on-board {\it XMM-Newton}.  We find that this transiting event, which lasts for roughly one day, comprises a mildly-ionized cloud of gas with a line-of-sight column density of $N_{\rm H} \sim 5 \times 10^{22}$\,cm$^{-2}$ that occults $\sim$60\,per cent of the continuum source.  We infer the likely location of this cloud to be the inner broad-line region.  In this paper, we focus on the X-ray variability properties of NGC\,3227, exploring time series, hardness ratios, primary spectral variations, the energy-dependent rms variability, the PSD, and Fourier time lags.

\section{Observations and data reduction} \label{sec:obs_and_data_reduction}

NGC\,3227 was observed six times by {\it XMM-Newton} \citep{Jansen01} over a month-long period from 2016-11-09 to 2016-12-09.  The six observations were co-ordinated with simultaneous {\it NuSTAR} \citep{Harrison13} observations.  There was also an additional, seventh {\it NuSTAR} observation roughly six weeks after the joint campaign on 2017-01-21, co-ordinated with a {\it Chandra} GTO observation.  An observation log is provided in Table~\ref{tab:obs_log}.  Below, we describe the data reduction procedures, which were performed using \textsc{HEAsoft}\footnote{\url{https://heasarc.nasa.gov/lheasoft/}} version 6.26 and version 18.0 of the {\it XMM-Newton} Scientific Analysis Software (\textsc{sas}\footnote{\url{http://xmm.esac.esa.int/sas/}}) package.

\subsection{XMM-Newton}

The total duration of the six {\it XMM-Newton} observations was $\sim$510\,ks, with individual observation lengths ranging from 74--92\,ks.  In this paper, we utilize X-ray data acquired with the European Photon Imaging Camera (EPIC)-pn \citep{Struder01}.  Meanwhile, simultaneous optical/UV coverage is provided by the co-aligned Optical Monitor (OM; \citealt{Mason01}).

\subsubsection{EPIC-pn}

The EPIC-pn observations were performed in `Small Window' mode using the medium optical filter.  The EPIC-pn events were processed using the \textsc{epproc}\footnote{\url{https://xmm-tools.cosmos.esa.int/external/sas/current/doc/epproc/index.html}} task within \textsc{sas}.  Circular regions 35\,arcsec radius were used to extract source events, including single- and double-pixel events (i.e. {\tt PATTERN} $\leq 4$).  Meanwhile, background events were extracted from larger circular regions away from the central source and avoiding the edges of the CCD.  Generally, the background level was relatively stable, although some periods of background flaring were apparent.  These typically occurred towards the beginning or end of an individual observation, usually lasting for just a few ks, and so were filtered out for the subsequent analysis.  The total net exposure after background filtering and, accounting for the detector dead time, was $\sim$310\,ks.  The average 0.3--10\,keV EPIC-pn count rate across all six observations was $\sim$8.6\,ct\,s$^{-1}$, corresponding to a time-averaged source flux of $F_{\rm 0.3-10} = 5.7 \times10^{-11}$\,erg\,cm$^{-2}$\,s$^{-1}$.  Meanwhile, the background count rate was found to be low, at $< 1$\,per cent of the source rate.  See Table~\ref{tab:obs_log} for a summary of the observations and their observed broad-band count rates and fluxes.

\begin{table*}
\centering
\begin{tabular}{l c c c c c c c c c}
\toprule
& & \multicolumn{4}{c}{{\it XMM-Newton} EPIC-pn} &  \multicolumn{4}{c}{{\it NuSTAR} FPMA+FPMB}\\
Obs. & Date & ID & Duration & Rate & Flux & ID & Duration & Rate & Flux \\
& & & (ks) & (ct\,s$^{-1}$) & ($\times 10^{-11}$\,erg\,cm$^{-2}$\,s$^{-1}$) & & (ks) & (ct\,s$^{-1}$) & ($\times 10^{-10}$\,erg\,cm$^{-2}$\,s$^{-1}$) \\
\midrule
1& 2016-11-09 & 0782520201 & 92 [49] & 7.62 & $4.56^{+0.02}_{-0.02}$ & 60202002002 & 98 [50] & 3.08 & $1.23^{+0.02}_{-0.02}$ \\
2& 2016-11-25 & 0782520301 & 74 [44] & 6.07 & $3.61^{+0.01}_{-0.02}$ & 60202002004 & 86 [42] & 2.62 & $1.03^{+0.02}_{-0.02}$ \\
3 & 2016-11-29 & 0782520401 & 84 [55] & 8.05 & $4.49^{+0.02}_{-0.01}$ & 60202002006 & 91 [40] & 2.92  & $1.15^{+0.02}_{-0.02}$ \\
4 & 2016-12-01 & 0782520501 & 87 [48] & 9.22 & $5.28^{+0.02}_{-0.02}$ & 60202002008 & 87 [42] & 3.42 & $1.33^{+0.02}_{-0.02}$ \\
5 & 2016-12-05 & 0782520601 & 87 [60] & 11.67 & $6.22^{+0.02}_{-0.02}$ & 60202002010 & 87 [41] & 3.59 & $1.38^{+0.02}_{-0.02}$ \\
6 & 2016-12-09 & 0782520701 & 88 [53] & 8.43 & $5.24^{+0.02}_{-0.01}$ & 60202002012 & 86 [39] & 3.37 & $1.29^{+0.02}_{-0.02}$ \\
7 & 2017-01-21 & --- & --- & --- & --- & 60202002014 & 95 [48] & 4.43 & $1.80^{+0.02}_{-0.02}$ \\
\bottomrule
\end{tabular}
\caption{Observation log of the {\it XMM-Newton} EPIC-pn and {\it NuSTAR} observations of NGC\,3227.  Net exposure times after filtering and accounting for `dead time' are provided in parentheses after the total durations.  The broad-band EPIC-pn and FPM count rates and fluxes are quoted from 0.3--10 and 3--78\,keV, respectively.}
\label{tab:obs_log}
\end{table*}

\subsubsection{Optical/UV Monitor}

The OM was operated in ``imaging'' mode using the UVW1 filter, which has a peak effective wavelength of 2\,910\,\AA.  A series of images was acquired in each observation for the purpose of UV monitoring.  A total of 95 images were acquired across the six {\it XMM-Newton} observations with a total OM exposure time of 375\,ks (typically $\sim$3--4\,ks per individual exposure).  All data were processed using the \textsc{omichain}\footnote{\url{http://xmm.esac.esa.int/sas/current/doc/omichain/}} task within \textsc{sas}.  This takes into account all calibration requirements and performs aperture photometry on the list of detected sources, providing count rates that are corrected for dead time and coincidence losses.

\subsection{NuSTAR}

{\it NuSTAR} is comprised of two Focal Plane Modules (FPMs): FPMA and FPMB, providing continuous coverage over a broad bandpass from 3--78\,keV.  The seven {\it NuSTAR} observations of NGC\,3227 covered a total duration of $\sim$300\,ks, with a typical duration of $\sim$40--50\,ks per observation.  To extract spectral products, we used the \textsc{nupipeline} and \textsc{nuproducts} scripts within \textsc{HEAsoft}, using the latest version of the calibration database (v20180419).  These were cleaned by applying standard screening criteria, such as filtering out passages through the South Atlantic Anomaly.  For each FPM, spectral products and light curves were extracted from circular source regions with a radius of 70\,arcsec.  Meanwhile, background products were extracted using 75\,arcsec circular regions separate from the source and away from the edges of the detector.  The time-averaged FPMA+FPMB 3--78\,keV background-subtracted count rate was $\sim$3.36\,ct\,s$^{-1}$, corresponding to an average observed broad-band flux of $F_{\rm 3-78} = 1.28 \times 10^{-10}$\,erg\,cm$^{-2}$\,s$^{-1}$.  See Table~\ref{tab:obs_log} for a summary of the observations.

\section{Results} \label{sec:results}

Here, we detail our main results from an analysis of the variability of NGC\,3227.  All data were fitted in \textsc{xspec} v.12.9.1 \citep{Arnaud96}.  All errors are quoted at the 90\,per cent confidence level, unless otherwise stated.

\subsection{Light curves} \label{sec:lightcurves}

\begin{figure*}
\begin{center}
\rotatebox{0}{\includegraphics[width=17.8cm]{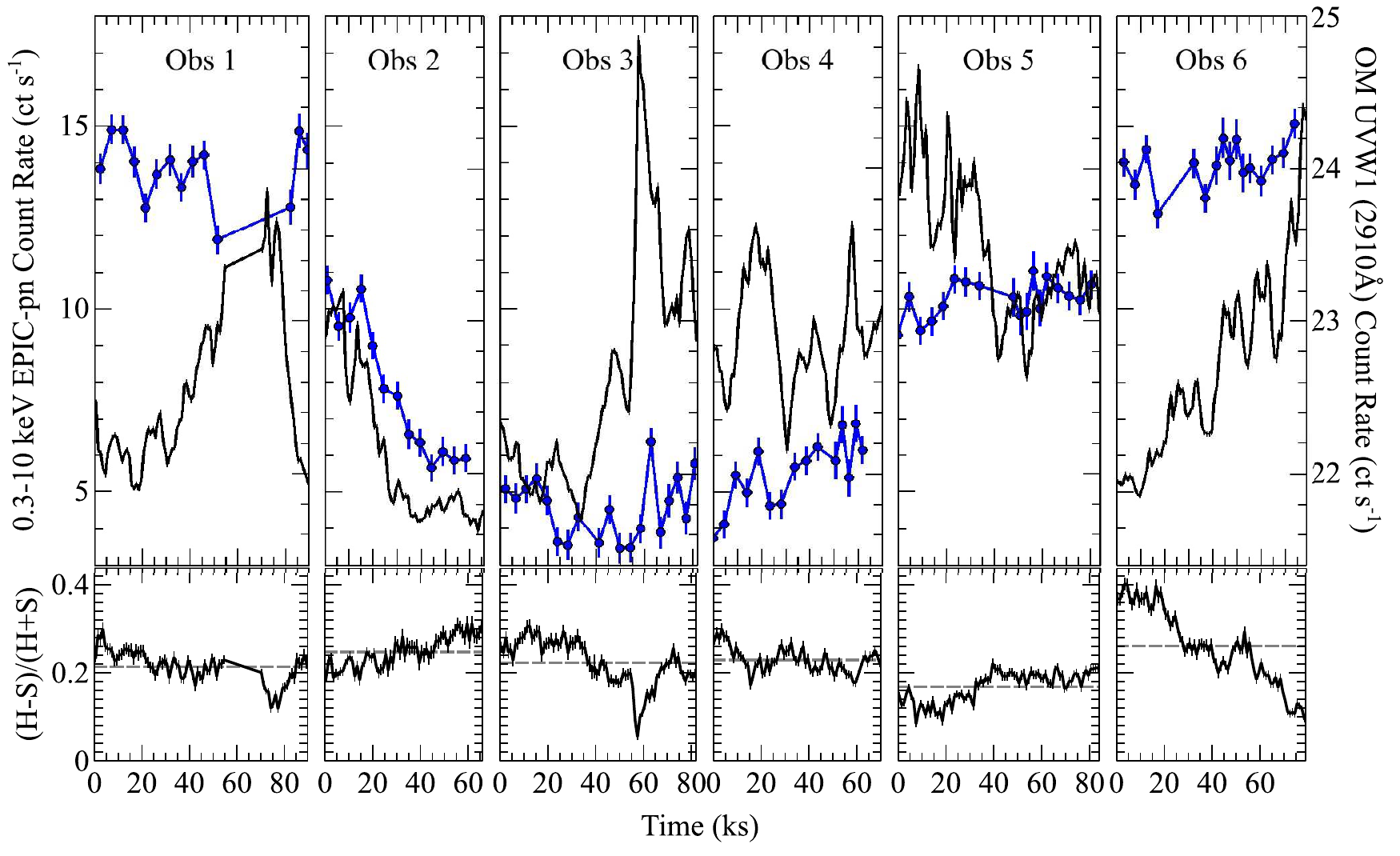}}
\end{center}
\vspace{-15pt}
\caption{Upper panel: the concatenated {\it XMM-Newton} EPIC-pn light curve of NGC\,3227 obtained from all six observations obtained in 2016.  The 0.3--10\,keV EPIC-pn X-ray light curve is shown in black, while simultaneous UV monitoring data acquired with the OM are shown in blue.  Lower panel: the evolution of the normalized hardness ratio in 1\,ks bins, defined as $(H-S)/(H+S)$, where the hard band, $H$, is 1--10\,keV and soft band, $S$, is 0.3--1\,keV.  The dashed horizontal lines represent the mean hardness ratio in each observation.}
\label{fig:pn_om_lc}
\end{figure*}

In this section, we show the {\it XMM-Newton} and {\it NuSTAR} light curves of NGC\,3227.  Fig~\ref{fig:pn_om_lc} (upper panel) shows the concatenated, background-subtracted {\it XMM-Newton} EPIC-pn light curve from 0.3--10\,keV in 1\,ks bins.  This is corrected for exposure losses and any telemetry drop-outs are interpolated over, where necessary.  Strong variability is clearly visible with the count rate varying by more than a factor of four over the course of the campaign.  The variability is observed to be rapid, even on within-orbit time-scales, with the source flux routinely doubling in just tens of ks.  We can quantify the intrinsic source variance by calculating the `excess variance' (\citealt{Nandra97}; \citealt{Edelson02}; \citealt{Vaughan03}), which takes into account the measurement uncertainties that also contribute to the total observed variance.  The excess variance is defined as $\sigma^{2}_{\rm XS} = S^{2} - \overline{\sigma^{2}_{\rm err}}$, where $S^{2} = \frac{1}{N-1}\sum\limits^{N}_{i=1}(x_{\rm i} - \overline{x})^{2}$ is the sample variance and $\overline{\sigma^{2}_{\rm err}} = \frac{1}{N}\sum\limits^{N}_{i=1}\sigma^{2}_{\rm err, i}$ is the mean square error.  The observed value is represented by $x_{\rm i}$, its arithmetic mean by $\overline{x}$ and the uncertainty on each individual measurement by $\sigma_{\rm err}$.  From this, we can calculate the fractional root mean square variability, $\label{eq:f_var} F_{\rm var} = \sqrt{\frac{\sigma^{2}_{\rm XS}}{\overline{x}^{2}}}$, which we can express as a percentage.  Note that this statistic is dependent upon the binning time-scale, which, in this case, is 1\,ks.  In the case of our broad-band 0.3--10\,keV X-ray light curve, the fractional variability is high at 33\,per cent over the course of the campaign.  In terms of within-observation variability, the fractional variability remains high, but varies across the six observations with values of 27, 34, 40, 16, 16, and 30\,per cent, respectively.  This can be visualized in Fig.~\ref{fig:pn_om_lc}, where NGC\,3227 is more variable in some observations than others.

Superimposed on the X-ray light curve in Fig.~\ref{fig:pn_om_lc} is the OM UVW1 ($2\,910$\,\AA) light curve (blue).  Each datapoint represents a single OM exposure.  The average observed corrected UV count rate is 22.9\,ct\,s$^{-1}$.  We can obtain a rough estimate on the flux in this band using the conversion factors calculated by the \textsc{sas} team \footnote{\url{http://www.cosmos.esa.int/web/xmm-newton/sas-watchout-uvflux}}.  This provides us with an estimated 2\,910\,\AA\ average flux of $1.1 \times 10^{-14}$\,erg\,cm$^{-2}$\,s$^{-1}$\,\AA$^{-1}$.  It is clear from the plot that the UV data are highly variable, even within the time-scale of a single observation ($<$\,day) --- particularly during obs\,2.  Quantitatively, the fractional variability is measured to be $\sim$4\,per cent (on time-scales of $\sim$3--4\,ks).  There is no clear long-term correlation between the UV data and the X-rays although a smooth transition is observed in the UV light curve as it appears to go through a pronounced dip in the middle of the campaign.  Curiously, some short-term, quasi-instantaneous correlated variability is observed --- most notably the sharp emission flare during obs\,3 ($\sim$215\,ks through the campaign), which is prominent in both the X-ray and UV light curves.  In terms of whether these UV variations could be energetically reproduced via X-ray reprocessing, we note that the observed X-ray variations are $\sim$25 times larger than those in the UV band, while the X-ray luminosity is $\sim$3 times weaker (at 1\,keV).  As such, larger X-ray variations would be sufficient to drive the observed smaller variations in the higher-luminosity UV band.

In the lower panel of Fig.~\ref{fig:pn_om_lc}, we show the evolution of the hardness ratio of NGC\,3227 across the course of the campaign.  We use the definition of the fractional hardness ratio: $HR = (H-S)/(H+S)$, where $H$ is the hard band and $S$ is the soft band (see \citealt{Park06} for further discussion).  In this instance, $H$ covers 1--10\,keV while $S$ covers the 0.3--1\,keV band.  Variability in the hardness ratio of the source is clearly visible.  Most notably, the strong flare in obs\,3 is accompanied by a sharp drop in the hardness ratio, indicating that the flare dominates in the soft band.  Additionally, significant evolution of the hardness ratio can be observed in obs\,6, as the source gradually softens as the source flux brightens over the $\sim$80\,ks period. 

\begin{figure}
\begin{center}
\rotatebox{0}{\includegraphics[width=8.4cm]{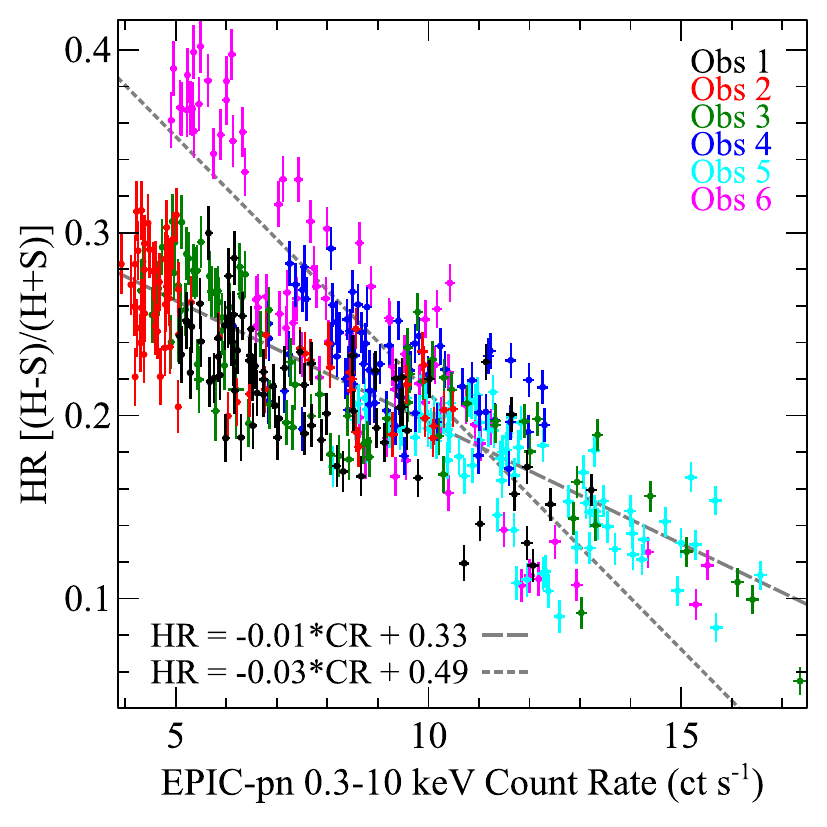}}
\end{center}
\vspace{-15pt}
\caption{The {\it XMM-Newton} EPIC-pn hardness ratio [$(H-S)/(H+S)$] plotted against the broad-band 0.3--10\,keV flux in 1\,ks bins.  The hard, $H$, and soft, $S$, bands are 1--10 and 0.3--1\,keV, respectively.  The six observations are shown in different colours: chronologically, black, red, green, blue, cyan, magenta, respectively.  The two grey lines show simple linear models fitted to the first five observations (dashed) and, separately, the sixth observation (dotted).}
\label{fig:pn_hr_vs_flux}
\end{figure}

In Fig.~\ref{fig:pn_hr_vs_flux}, we plot the hardness ratios against the broad-band 0.3--10\,keV count rate in 456 1\,ks bins across all six observations.  A strong correlation is observed (Pearson correlation coefficient: $r = -0.765$; $p < 10^{-5}$) implying that the source becomes softer when the flux is higher.  Fitting a simple linear model to the data returns values of the slope, $a = -0.016 \pm 0.001$ and the offset, $b = 0.356 \pm 0.002$ ($\Delta \chi^{2}/{\rm d.o.f.} = 4\,229/454$).  However, it is clear from the plot that a different mode of variability was present during obs\,6 (magenta), also apparent in the marked evolution of the hardness ratio in Fig.~\ref{fig:pn_om_lc}.  This difference in the shape of the spectrum in obs\,6 formed the motivation behind the analysis in \citet{Turner18}, where we show that NGC\,3227 appeared to undergo a strong occultation event by a cloud of ionized gas manifesting itself in a strengthening of the depth of the UTA in the high-resolution RGS data.  As such, we fitted obs\,1--5 and obs\,6 separately finding significant differences in the hardness-ratio slopes.  In the former case, $a = -0.013 \pm 0.001$ and $b = 0.329 \pm 0.002$ ($\Delta \chi^{2}/{\rm d.o.f.} = 2\,212 / 375$).  Meanwhile, in the latter case, $a = -0.028 \pm 0.001$ and $b = 0.493 \pm 0.005$ ($\Delta \chi^{2}/{\rm d.o.f.} = 848/77$).  These two fits are shown in Fig.~\ref{fig:pn_hr_vs_flux}.

\begin{figure*}
\begin{center}
\rotatebox{0}{\includegraphics[width=17.8cm]{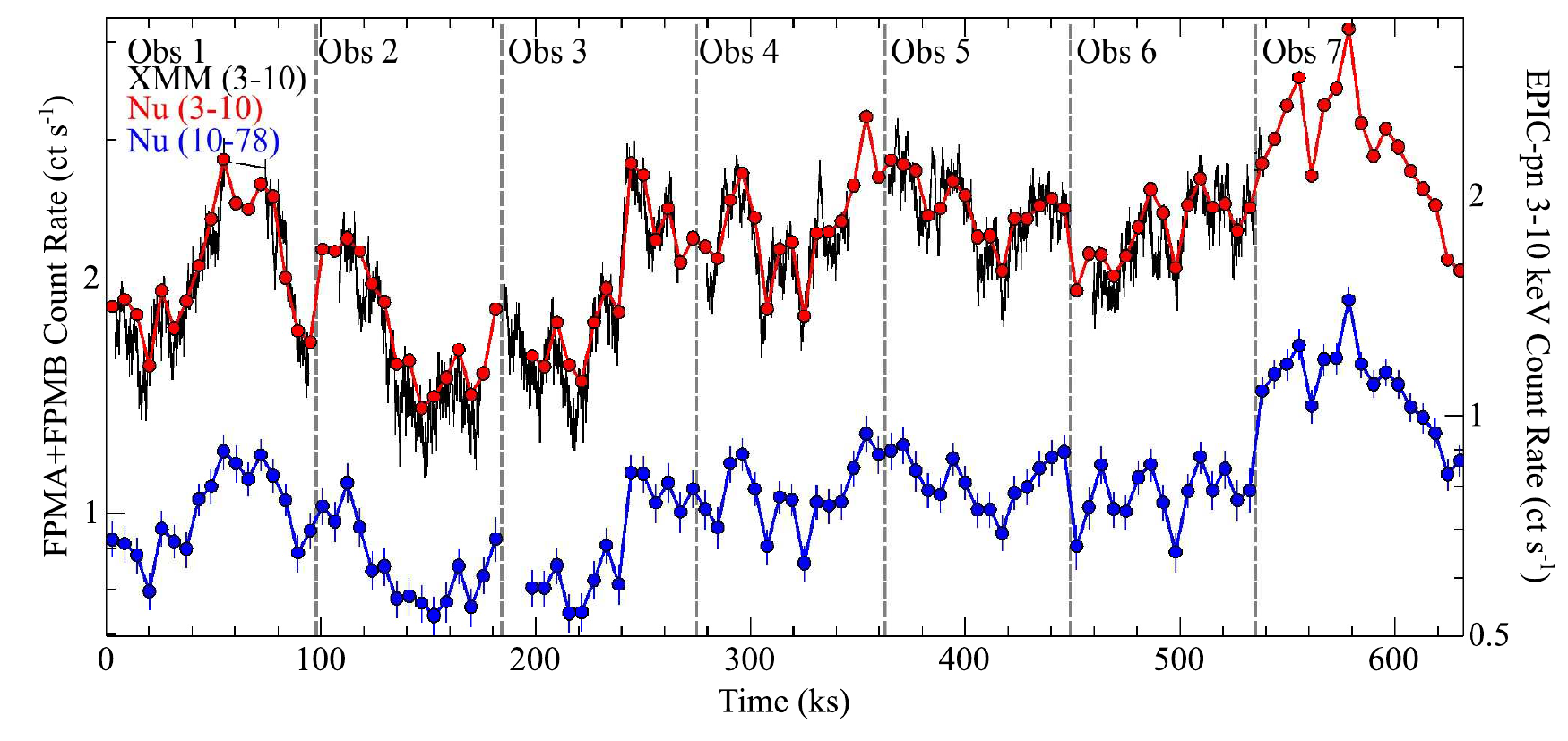}}
\end{center}
\vspace{-15pt}
\caption{The concatenated {\it NuSTAR} FPMA+FPMB light curve in orbital bins (96.8\,min).  The overlapping 3--10\,keV {\it NuSTAR} and {\it XMM-Newton} bands are shown in red and black (latterly: 500\,s bins), respectively.  Meanwhile, the harder 10--78\,keV {\it NuSTAR} light curve is shown in blue.  Note the log scale on the $y$-axis.}
\label{fig:nustar_xmm_lc}
\end{figure*}

We now include the {\it NuSTAR} data by creating a concatenated light curve from all seven observations.  We combined data from the two FPMs and used orbital bins (96.8\,min).  This is shown in Fig.~\ref{fig:nustar_xmm_lc}.  We firstly plotted the light curve in the 3--10\,keV band, which overlaps with the {\it XMM-Newton} EPIC-pn bandpass, which we superimpose using 500\,s bins.  It is clear from the plot that the {\it NuSTAR} and {\it XMM-Newton} light curves show near-identical behaviour and are well correlated.  We note that the start/stop times of the {\it NuSTAR} observations are sometimes offset slightly from {\it XMM-Newton}.  We also show the much harder, 10--78\,keV {\it NuSTAR} light curve, which shows similar behaviour.  Although the absolute strength of the variability is suppressed, the strength of the fractional variability is similar ($F_{\rm var} = 20$\,per cent in the 10--78\,keV band versus $F_{\rm var} = 23$\,per cent in the 3--10\,keV band).

\subsection{Spectral decomposition} \label{sec:spectral_decomposition}

\begin{figure}
\begin{center}
\rotatebox{0}{\includegraphics[width=8.4cm]{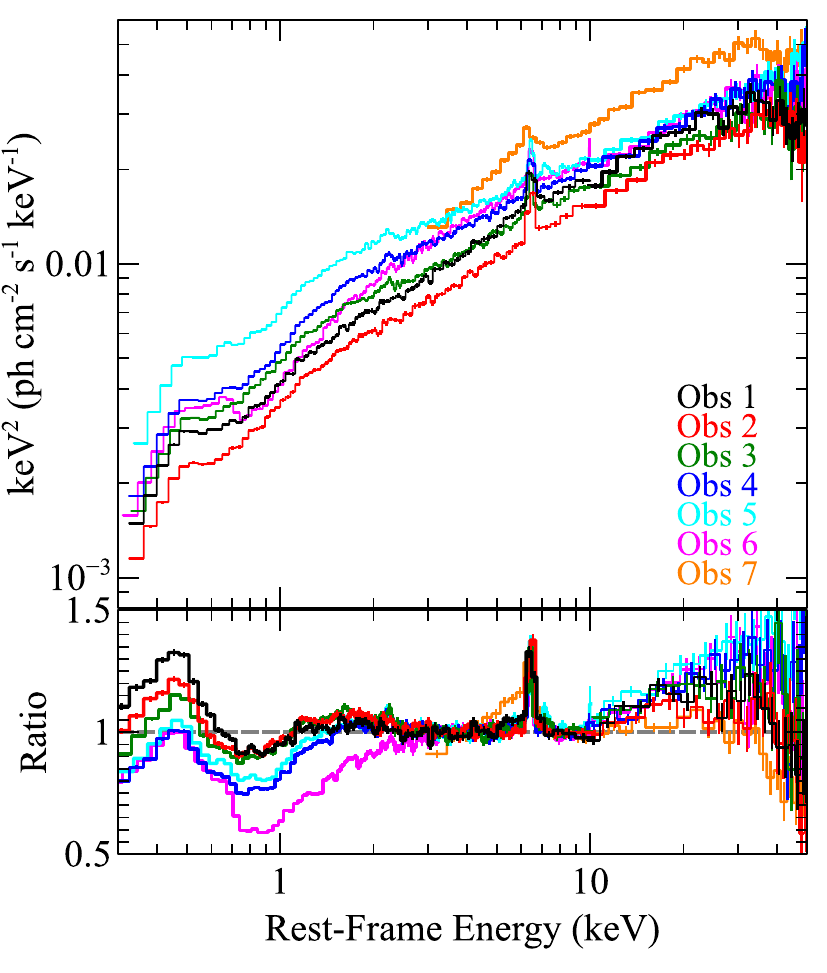}}
\end{center}
\vspace{-15pt}
\caption{Upper panel: all six {\it XMM-Newton} EPIC-pn and all seven {\it NuSTAR} FPMA+FPMB spectra ``fluxed'' against a flat power law with $\Gamma = 0$.  The EPIC-pn spectra are shown from 0.3--10\,keV while the {\it NuSTAR} spectra are shown $> 10$\,keV only for clarity, except for obs\,7, which is shown from 3--50\,keV.  The seven observations are shown, chronologically, in black, red, green, blue, cyan, magenta, and orange, respectively.  Lower panel: the ratio to an absorbed power law fitted in a continuum-dominated band from 3.0--5.5 and 7.5--10\,keV and extrapolated over the entire bandpass.  See Section~\ref{sec:spectral_decomposition} for details.}
\label{fig:pn_nustar_eeuf_ratio}
\end{figure}

Here, we investigate the spectral variability NGC\,3227 across the observing campaign.  In Fig.~\ref{fig:pn_nustar_eeuf_ratio} we show all six {\it XMM-Newton} EPIC-pn and all seven {\it NuSTAR} spectra.  For the {\it NuSTAR} spectra, we combined the FPMA and FPMB data in our plots for clarity.  For further clarity, we only show the {\it NuSTAR} spectra $> 10$\,keV (below which there is overlap with the EPIC-pn band) and $< 50$\,keV as the spectra become noisy at the highest energies.  The spectra are ``fluxed'' against a power law with a flat photon index (i.e. $\Gamma = 0$).  It is clear that all the spectra are hard with some modest variations in flux, with the bulk of the variability occurring at lower energies while the spectra tend to converge at the highest energies.

To visualize the clearest emission and absorption components, we fitted the spectra with a simple power law, absorbed by a neutral Galactic column of $N_{\rm H} = 2.13 \times 10^{20}$\,cm$^{-2}$, in which the measurements of \citet{Kalberla05} are modified by taking into account the additional effect of molecular hydrogen (see \citealt{Willingale13}).  The Galactic column was modelled with the \textsc{tbabs} component within \textsc{xspec} \citep{WilmsAllenMcCray00}, using the appropriate photoionization cross-sections \citep{Verner96}.  We fitted the spectrum over continuum-dominated bands that are usually free from obvious emission/absorption features: 3.0--5.5 and 7.5--10\,keV.  We then extrapolated the fit over the entire 0.3--50\,keV bandpass, allowing the photon index and normalization to vary between observations.  The spectra are all observed to be hard with the photon index measurements falling in the range:  1.4--1.7.  At energies $< 2$\,keV, it is clear that significant absorption and emission components are present.  In particular, strong absorption is present in obs\,6 (magenta), which is most prominent at $\sim$0.7--0.9\,keV and is due to absorption by the UTA.  This is the signature of the occultation event described in \citet{Turner18}.  Meanwhile, evidence of emission is apparent at higher energies, most notably in the form of a strong emission line at 6.4\,keV due to near-neutral Fe\,K$\alpha$.  Some additional emission with respect to the simple power law is also visible $> 10$\,keV, most likely arising from modest Compton reflection.

\begin{figure}
\begin{center}
\rotatebox{0}{\includegraphics[width=8.4cm]{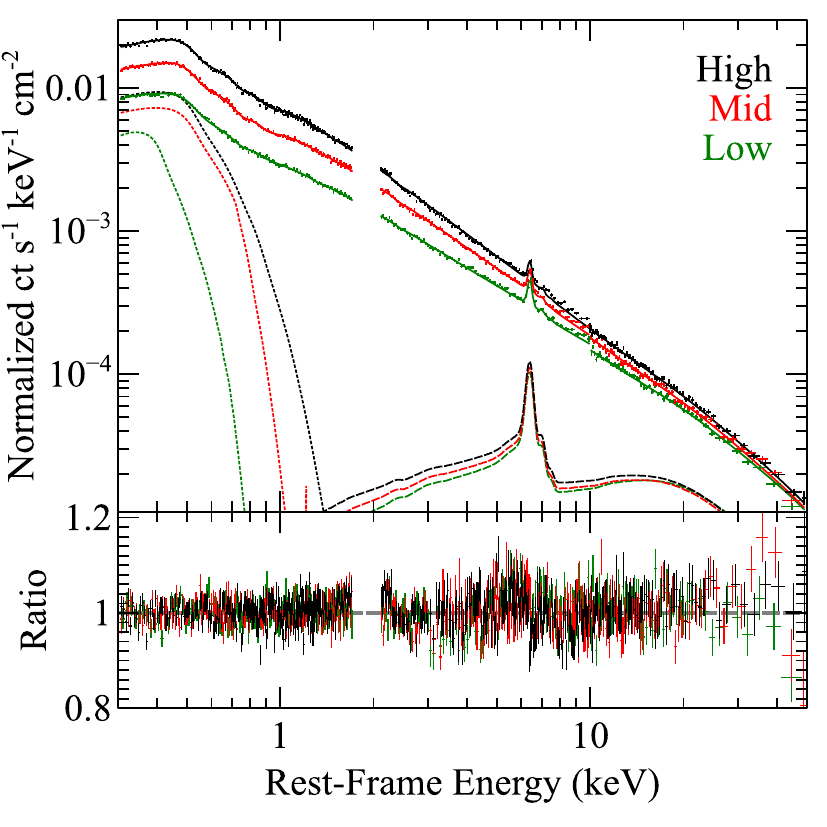}}
\end{center}
\vspace{-15pt}
\caption{Upper panel: the high- (black), mid- (red), and low- (green) flux spectra of NGC\,3227, obtained by splitting the data into three flux-resolved segments.  The data are fitted with the model described in Section~\ref{sec:spectral_decomposition}.  The contributions of the blackbody and neutral reflection components are shown by the dotted and dashed lines, respectively.  Lower panel: the ratio of the residuals to the model.  Note that all data are binned up and that the {\it NuSTAR} data are only plotted $>10$\,keV for clarity.  Meanwhile, the EPIC-pn data are ignored from 1.7--2.1\,keV due to calibration uncertainties around the Si\,K edge in the detector.  See Section~\ref{sec:spectral_decomposition} for details.}
\label{fig:high_mid_low_flux_spectra}
\end{figure}

To investigate the spectral variability, we created high-, medium-, and low-flux broad-band spectra.  Here, we split the {\it XMM-Newton} EPIC-pn  and {\it NuSTAR} FPM data into three flux-resolved groups.  To achieve this, we firstly defined good time intervals (GTIs) that were common to both satellites, merging them using the \textsc{mgtime} task.  We then used the combined GTI to create a broad-band light curve across the whole campaign in 50\,s bins, yielding $\sim$230\,ks worth of common data.  We then split this into three separate flux-resolved groups with identical exposure times\footnote{We note that applying cuts according to exposure time can lead to biases - i.e. greater statistical weighting may be applied to the higher-flux spectra due to the larger number of counts.  As such, we also created flux-resolved spectra applying cuts such that the total number of counts in each flux-slice was identical.  In this case, cuts were applied at $>10.85$, $8.39-10.85$ and $<8.39$\,ct\,s$^{-1}$ with respective exposure times of 51.5, 67.7 and 106.2\,ks.  However, the results were consistent with those described in Section~\ref{sec:spectral_decomposition}.  Additionally, we also tried a simpler approach, by using the highest-flux observation (obs\,5) and the lowest-flux observation (obs\,2) to create difference spectra.  Again, the results were consistent.}; i.e. $\sim$75\,ks each of high-, mid-, and low-flux data.  In the 0.3--10\,keV EPIC-pn case (as per Fig.~\ref{fig:pn_om_lc}), the cuts were applied at $> 9.97$, $6.90-9.97$, and $<6.90$\,ct\,s$^{-1}$.  This allowed us to create high-, mid-, and low-flux GTIs which we then applied to the respective processing pipelines to create high-, mid- and low-flux spectra for each observation.  The flux-resolved {\it XMM-Newton} and {\it NuSTAR} spectra were then combined using the \textsc{mathpha} task.  Meanwhile, response files were created by averaging across the six observations using the \textsc{addrmf} and \textsc{addarf} tasks, weighting them by the appropriate exposure times.  This resulted in three {\it XMM-Newton} EPIC-pn and {\it NuSTAR} FPM spectra in three separate flux bands, which we binned such that there were $>25$\,ct\,bin$^{-1}$.

These spectra were then fitted within \textsc{xspec} with a broad-band model based on the spectral fitting described in \citet{Turner18}.  The spectra are shown in Fig.~\ref{fig:high_mid_low_flux_spectra} and are found to have observed broad-band 0.3--50\,keV fluxes, from high-to-low flux, of: $1.36^{+0.02}_{-0.01} \times 10^{-10}$, $1.14^{+0.01}_{-0.02} \times 10^{-10}$, and $9.51^{+0.02}_{-0.02} \times 10^{-11}$\,erg\,cm$^{-2}$\,s$^{-1}$.  NGC\,3227 is clearly more variable at lower energies with the X-rays varying by up to $\sim$80\,per cent $< 10$\,keV compared to $< 20$\,per cent $> 10$\,keV.  Note that, while we fitted the {\it NuSTAR} FPMA and FPMB spectra separately, we combined them for the purposes of plotting.  Additionally, while the {\it NuSTAR} data were fitted $> 3$\,keV, we only show them $> 10$\,keV for clarity.  The model consists of: (i) a primary power-law continuum, (ii) a \textsc{pexmon} component \citep{Nandra07} to model neutral reflection and associated Fe\,K$\alpha$ emission, (iii) a high-energy cut-off, (iv) a blackbody component to model the soft excess, and (v) additional multiplicative components to account for the warm absorber.  The warm absorber in NGC\,3227 consists of three separate zones, which we modelled with version 2.41 of the \textsc{xstar} photoionization code \citep{KallmanBautista01, Kallman04}.  Each zone is characterized by its column density, $N_{\rm H}$, ionization parameter, $\xi$\footnote{The ionization parameter is defined as $\xi = L_{\rm ion}/n_{\rm e}r^{2}$ and has units erg\,cm\,s$^{-1}$, where $L_{\rm ion}$ is the ionizing luminosity from 1--1\,000\,Rydberg in units of erg\,cm$^{-2}$\,s$^{-1}$, $n_{\rm e}$ is the gas density in cm$^{-3}$ and $r$ is the distance of the absorbing gas from the ionizing source in cm.} and outflow velocity, $v_{\rm out}$.  The best-fitting values of the ionization parameter and outflow velocity across the three zones (zones 1, 2, 3), respectively, are as follows: log\,$\xi = -0.7$, $1.4$, and $2.9$ and $v_{\rm out} = 100$, $250$, and $1\,300$\,km\,s$^{-1}$.  These come from our analysis of the RGS data (see \citealt{Turner18}) and were fixed in the fit and tied across all three spectra, while the column densities were allowed to vary.  We also allowed the respective normalizations and photon indices to vary between spectra.

The best-fitting photon indices (tied between the primary power law and the reflection component) were found to be $\Gamma = 1.72 \pm 0.01$, $1.60 \pm 0.01$ and $1.40 \pm 0.02$ for the high-, medium- and low-flux spectra, respectively.  This is consistent with the steeper-when-brighter behaviour described in Section~\ref{sec:lightcurves}.  The blackbody component has a best-fitting temperature of $kT = 0.09 \pm 0.02$\,keV while the e-folding energy of the high-energy cutoff is found to be $E_{\rm cut} = 300 \pm 80$\,keV.  Meanwhile, the best-fitting values of the column densities suggest weak variability in the strength of the warm absorber (given the measurement uncertainties) across all three flux-selected spectra (high-, medium-, low-flux, respectively), with the following values: zone 1:  $N_{\rm H} = 1.46 ^{+0.12}_{-0.09} \times 10^{21}$, $9.69^{+0.15}_{-0.19} \times 10^{20}$ and $2.74^{+0.27}_{-0.20} \times 10^{21}$\,cm$^{-2}$; zone 2: $N_{\rm H} = 1.43^{+0.44}_{-0.46} \times 10^{21}$, $1.29^{+0.98}_{-0.59} \times 10^{21}$ and $2.45^{+0.78}_{-0.88} \times 10^{21}$\,cm$^{-2}$; zone 3: $N_{\rm H} = 1.26^{+1.41}_{-1.36} \times 10^{21}$, $8.01^{+2.22}_{-2.46} \times 10^{20}$ and $7.42^{+2.41}_{-1.40} \times 10^{20}$\,cm$^{-2}$.  The overall fit to the three spectra is good: $\chi^{2}/{\rm d.o.f.} = 7\,839/7\,283$.

An alternative test was performed to search for variability of the warm absorber in terms of its ionization parameter instead of column density.  Such behaviour has been observed in time-resolved RGS analysis of other AGN (e.g. NGC\,4051; \citealt{Krongold07}) where log\,$\xi$ is seen to correlate with the luminosity of the irradiating power law on short time-scales.  One might expect to observe such behaviour if the warm absorber is ionized by the central AGN.  In this instance, we fixed the column densities at their best-fitting values from our RGS analysis in \citet{Turner18}.  These are as follows: $N_{\rm H} = 2.1 \times 10^{21}$, $8.3 \times 10^{20}$, and $4.4 \times 10^{21}$\,cm$^{-2}$ for the three zones, respectively.  In this case, we find no evidence of flux-dependent variations in the ionization parameter with the best-fitting values remaining consistent within the measurement uncertainties across all three spectra.  Moreover, the fit is worse by $\Delta \chi^{2} = 43$ than in the case where the column densities are allowed to vary.

\begin{table*}
\centering
\begin{tabular}{l c c c c c c c c c}
\toprule
\multirow{3}{*}{Spectrum} & Zone\,1 & Zone\,2 & Zone\,3 & \multicolumn{2}{c}{Power Law} & \textsc{bbody} & \textsc{pexmon} & \multicolumn{2}{c}{Fe\,K$\alpha$}  \\
& $N_{\rm H}$ & $N_{\rm H}$ & $N_{\rm H}$ & $\Gamma$ & Flux & Flux & Flux & EW & Flux  \\
&($\times$10$^{21}$) & ($\times$10$^{21}$) & ($\times$10$^{21}$) & & ($\times$10$^{-10}$) & ($\times$10$^{-13}$) & ($\times$10$^{-11}$) & (keV) & ($\times$10$^{-13}$) \\
\midrule
High & $1.46^{+0.12}_{-0.09}$ & $1.43^{+0.44}_{-0.46}$ & $1.26^{+1.41}_{-1.36}$ & $1.72 \pm 0.01$ & $1.50^{+0.03}_{-0.02}$ & $3.31^{+0.06}_{-0.06}$ & $1.58^{+0.08}_{-0.09}$ & $0.10 \pm 0.01$ & $4.83^{+0.47}_{-0.46}$ \\
Mid & $0.97^{+0.02}_{-0.02}$ & $1.29^{+0.98}_{-0.59}$ & $0.80^{+0.22}_{-0.25}$ & $1.60 \pm 0.01$ & $1.25^{+0.02}_{-0.02}$ & $3.03^{+0.04}_{-0.05}$ & $1.38^{+0.10}_{-0.09}$ & $0.13 \pm 0.01$ & $4.46^{+0.45}_{-0.43}$  \\
Low & $2.74^{+0.27}_{-0.20}$ & $2.45^{+0.78}_{-0.88}$ & $0.74^{+0.24}_{-0.14}$ & $1.40 \pm 0.02$ & $0.91^{+0.04}_{-0.04}$ & $2.56^{+0.05}_{-0.04}$ & $1.24^{+0.10}_{-0.10}$ & $0.16^{+0.02}_{-0.01}$ & $4.46^{+0.45}_{-0.43}$ \\
\bottomrule
\end{tabular}
\caption{Table showing the best-fitting parameters of the variable components fitted to the flux-resolved spectra described in Section~\ref{sec:spectral_decomposition}.  All spectra are fitted from 0.3--50\,keV.  All column densities are given in units of cm$^{-2}$ and all fluxes are given in units of erg\,cm$^{-2}$\,s$^{-1}$.}
\label{tab:high_mid_low_parameters}
\end{table*}

We also performed a test to search for any variability of the reflection component by measuring its (unabsorbed) flux across the three flux-selected spectra.  We find that the broad-band flux of the \textsc{pexmon} component is $1.58^{+0.08}_{-0.09} \times 10^{-11}$, $1.38^{+0.10}_{-0.09} \times 10^{-11}$, and $1.24^{+0.10}_{-0.10} \times 10^{-11}$\,erg\,cm$^{-2}$\,s$^{-1}$ in the high-, mid-, and low-flux spectra, respectively.  This corresponds to modest variations of around $\sim$20\,per cent, which is significantly weaker than the variability of the primary power law and associated soft excess (see the values reported in Table~\ref{tab:high_mid_low_parameters}).  Its overall contribution to the model is also found to be of moderate strength, at $\sim$10\,per cent from 0.3--50\,keV (and $\sim$20\,per cent $> 10$\,keV).

Finally, we tested to see whether there was any variability in the component of Fe\,K$\alpha$ emission at $\sim$6.4\,keV.  Here, we took the baseline model described above but replaced the \textsc{pexmon} component with a \textsc{pexrav} component \citep{MagdziarzZdziarski95}, which only models the reflection continuum and not the associated emission line.  We then parametrized the emission line at 6.4\,keV independently with a Gaussian.  We find that the centroid energy, $E_{\rm c}$, and width, $\sigma$, of the line do not vary with flux with best-fitting values of $E_{\rm c} = 6.43 \pm 0.01$\,keV and $\sigma = 70 \pm 10$\,eV, while the equivalent width of the line is found to be, from high-, to mid-, to low-flux, respectively: EW = $0.10 \pm 0.01$, $0.13 \pm 0.01$, and $0.16^{+0.02}_{-0.01}$\,keV.  Meanwhile, the normalization of this component is $N = 4.71^{+0.46}_{-0.45} \times 10^{-5}$, $4.35^{+0.43}_{-0.42} \times 10^{-5}$, and $4.28^{+0.40}_{-0.39} \times 10^{-5}$\,ph\,cm$^{-2}$\,s$^{-1}$ from high-to-low flux, corresponding to observed line fluxes of $F_{\rm line} = 4.83^{+0.47}_{-0.46} \times 10^{-13}$, $4.46^{+0.45}_{-0.43} \times 10^{-13}$, and $4.39^{+0.41}_{-0.40} \times 10^{-13}$\,erg\,cm$^{-2}$\,s$^{-1}$.  These modest variations in the best-fitting values of the normalization / flux of the line are roughly consistent with the $\sim$20\,per cent variations observed in the reflection component, although we do not detect independent variations in the strength of the emission line within the 90\,per cent confidence uncertainties quoted here.  As such, only mild variations in near-neutral reflection are observed on the time-scales probed here.  All of these parameters are listed in Table~\ref{tab:high_mid_low_parameters}.

\subsubsection{Difference spectra} \label{sec:difference_spectra}

\begin{figure}
\begin{center}
\rotatebox{0}{\includegraphics[width=8.4cm]{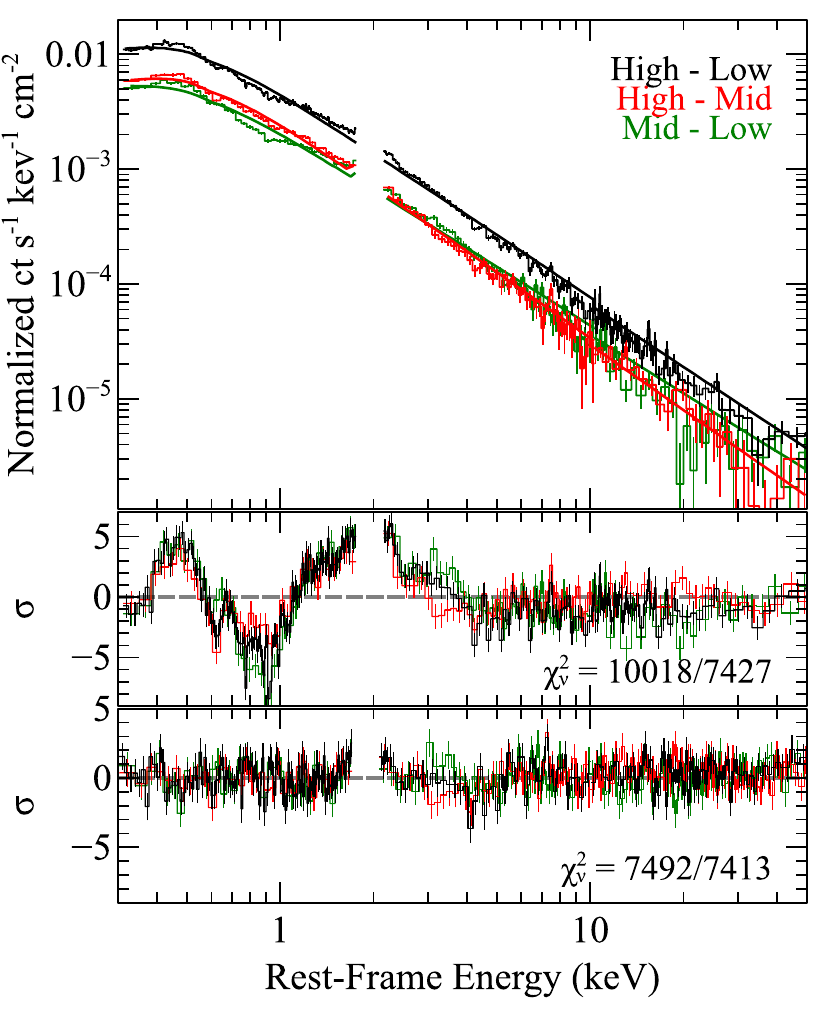}}
\end{center}
\vspace{-15pt}
\caption{Upper panel: the difference spectra of NGC\,3227, obtained by splitting the data into three flux-resolved segments.  The {\it XMM-Newton} EPIC-pn and {\it NuSTAR} FPMA+FPMB spectra are shown in three cases: high$-$low (black), high$-$mid (red), and mid$-$low (green).  Middle panel: the $\sigma$ residuals to a simple fit consisting of an absorbed power-law model.  Lower panel: the $\sigma$ residuals to a broad-band fit including three warm absorber zones.  Note that all data are binned up and that the {\it NuSTAR} data are only plotted $>10$\,keV for clarity.  Meanwhile, the EPIC-pn data are ignored from 1.7--2.1\,keV due to calibration uncertainties around the Si\,K edge in the detector.  See Section~\ref{sec:difference_spectra} for details.}
\label{fig:pn_nu_diff}
\end{figure}

To further investigate the spectral variability, we computed broad-band difference spectra, based on the flux-selected spectra defined above in Section~\ref{sec:spectral_decomposition}.  Three difference spectra were created by subtracting (i) low-flux data from high-flux data, (ii) mid-flux data from high-flux data and (iii) low-flux data from mid-flux data across the broad 0.3--50\,keV bandpass.  These were then fitted within \textsc{xspec} with the results shown in Fig.~\ref{fig:pn_nu_diff}.  Again, we fitted the {\it NuSTAR} FPMA and FPMB spectra separately, but combined them for the purposes of plotting.  Likewise, the {\it NuSTAR} data were fitted $> 3$\,keV, but are only shown $> 10$\,keV for clarity.  We applied a simple absorbed power-law model (\textsc{tbabs} $\times$ \textsc{pl}) to the spectra, tying the photon index, $\Gamma$, and column density, $N_{\rm H}$, between the {\it XMM-Newton} and {\it NUSTAR} spectra in each flux-resolved case.  We allowed the power-law normalizations to vary to account for variations in the cross-normalization.  The best-fitting power-law slopes are (i) $\Gamma = 1.81 \pm 0.01$, (ii) $\Gamma = 1.91 \pm 0.02$, and (iii) $\Gamma = 1.70 \pm 0.02$, while the neutral absorber column density lies in the range: $3-5 \times 10^{20}$\,cm$^{-2}$.  The overall fit statistic is $\chi^{2}/{\rm d.o.f.} = 10\,018 / 7\,427$ with clear residuals in the soft X-ray band.  Indeed, the fit is very poor $<3$\,keV: $\chi^{2}/{\rm d.o.f.} = 3\,579 / 1\,371$, although a visual examination of the deviations suggest that the spectral variability remains similar across all the flux ranges afforded by this observing campaign.

To account for the residuals in the soft band, we then included additional soft components, as required by broad-band spectral modelling (see \citealt{Turner18}).  We firstly included a blackbody component to model the soft excess.  Tying this component between all three difference spectra yielded a best-fitting temperature of $kT = 0.09 \pm 0.01$\,keV, with no requirement for the temperature to vary between spectra.  We then also included additional multiplicative components to account for the warm absorber (see Section~\ref{sec:spectral_decomposition}).  Again, the ionization parameters and outflow velocities were fixed in the fit and tied across all three spectra, while the column densities were allowed to vary.  The best-fitting values across all three spectra (high $-$ low, high $-$ mid, mid $-$ low, respectively) are as follows: zone 1:  $N_{\rm H} = 1.65 ^{+0.11}_{-0.08} \times 10^{21}$, $1.59^{+0.18}_{-0.16} \times 10^{21}$, and $1.74^{+0.22}_{-0.17} \times 10^{21}$\,cm$^{-2}$; zone 2: $N_{\rm H} = 2.88^{+0.49}_{-0.44} \times 10^{21}$, $1.14^{+0.95}_{-0.62} \times 10^{21}$, and $4.73^{+0.74}_{-0.94} \times 10^{21}$\,cm$^{-2}$; zone 3: $N_{\rm H} = 1.94^{+1.46}_{-1.27} \times 10^{21}$, $3.21^{+2.42}_{-2.69} \times 10^{21}$, and $1.32^{+2.93}_{-1.32} \times 10^{21}$\,cm$^{-2}$.  Meanwhile, the best-fitting photon indices of the power law are found to be $\Gamma = 2.10 \pm 0.02$, $2.13 \pm 0.04$, and $2.09 \pm 0.03$, respectively.  These components are found to significantly improve the fit, resulting in excellent joint broad-band fit overall: $\chi^{2}/{\rm d.o.f.} = 7\,492/7\,413$, with a large improvement in the soft band $< 3$\,keV: $\chi^{2}/{\rm d.o.f.} = 1\,449/1\,357$ (compared with $\chi^{2}/{\rm d.o.f.} = 3\,579 / 1\,371$ previously).  We note that the photon indices are steeper than those inferred from the time-averaged broad-band spectrum.  NGC\,3227, like other similar sources, exhibits typical softer-when-brighter behaviour, which can naturally lead to a steeper photon index in the difference spectrum than the average spectrum --- for example, were the power law to pivot, we would observe greater changes in flux at energies further away from the pivot point.  We also note that additional components are likely to contribute to the enhanced variability in the soft band.  For example, the spectrum requires an additional soft-band component such as a Comptonized disc blackbody, which can add to the variability at lower energies.  Additionally, any inter-orbital variations of the warm absorber may also enhance the low-energy variability at low frequencies. The residuals to this fit are shown in the lower panel of Fig.~\ref{fig:pn_nu_diff} and the best-fitting variable parameters are provided in Table~\ref{tab:diff_spectra_parameters}.  As before, we tested to see if variations in the ionization parameter could instead account for the spectral changes, but we again found that log\,$\xi$ remains constant within the measurement uncertainties across all three difference spectra.

\begin{table}
\centering
\begin{tabular}{l c c c}
\toprule
\multirow{2}{*}{Parameter} & \multicolumn{3}{c}{Difference Spectrum} \\
& High - Low & High - Mid & Mid - Low \\
\midrule
$\Gamma$ & $2.10 \pm 0.02$ & $2.13 \pm 0.04$ & $2.09 \pm 0.03$ \\
Zone\,1: $N_{\rm H}$ & $1.65^{+0.11}_{-0.08}$ & $1.59^{+0.18}_{-0.16}$ & $1.74^{+0.22}_{-0.17}$ \\
Zone\,2: $N_{\rm H}$ & $2.88^{+0.49}_{-0.44}$ & $1.14^{+0.95}_{-0.62}$ & $4.73^{+0.74}_{-0.94}$ \\
Zone\,3: $N_{\rm H}$ & $1.94^{+1.46}_{-1.27}$ & $3.21^{+2.42}_{-2.69}$ & $1.32^{+2.93}_{-1.32}$ \\
\bottomrule
\end{tabular}
\caption{The best-fitting variable parameters from the fits to the broad-band 0.3--50\,keV flux-resolved difference spectra described in Section~\ref{sec:difference_spectra}.  All column densities are given in units of $10^{21}$\,cm$^{-2}$.}
\label{tab:diff_spectra_parameters}
\end{table}

We did also create difference spectra by filtering on time as opposed to flux.  A simple broad-band difference spectrum was created by subtracting the lowest-flux obs\,2 EPIC-pn and FPM spectra from the highest-flux obs\,5 spectra.  We again applied the model described above finding that it provides a similarly good fit from 0.3--50\,keV.  The best-fitting values are largely consistent with those obtained from the flux-selected difference spectra; i.e. $\Gamma =  2.04 \pm 0.02$, $kT = 0.10 \pm 0.01$\,keV, and $N_{\rm H} = 1.54^{+0.09}_{-0.10} \times 10^{21}$, $3.96^{+0.67}_{-0.60} \times 10^{21}$, and $< 3.20 \times 10^{21}$\,cm$^{-2}$ for the three warm absorber zones, respectively.  The overall fit statistic is $\chi^{2}/{\rm d.o.f.} = 2\,280/2\,249$.

\subsection{The rms spectrum} \label{sec:rms_spectrum}

In this section, we compute the `rms spectrum', which is the rms amplitude of variability as a function of energy (e.g. \citealt{Vaughan03}).  We used the broad-band {\it XMM-Newton} EPIC-pn data, extracting light curves of equal segment-length in 20 energy bands.  These were equally spaced in log$(E)$ from 0.3--10\,keV.  The fractional excess variance, as given by $\sigma_{\rm xs}^2$ / mean$^{2}$, was then calculated for each of the 20 energy bands and averaged over each segment.  To calculate the fractional rms (or fractional variability, $F_{\rm var}$), we then took the square root of the excess variance and averaged this over all {\it XMM-Newton} observations from 2016.

\begin{figure}
\begin{center}
\rotatebox{0}{\includegraphics[width=8.4cm]{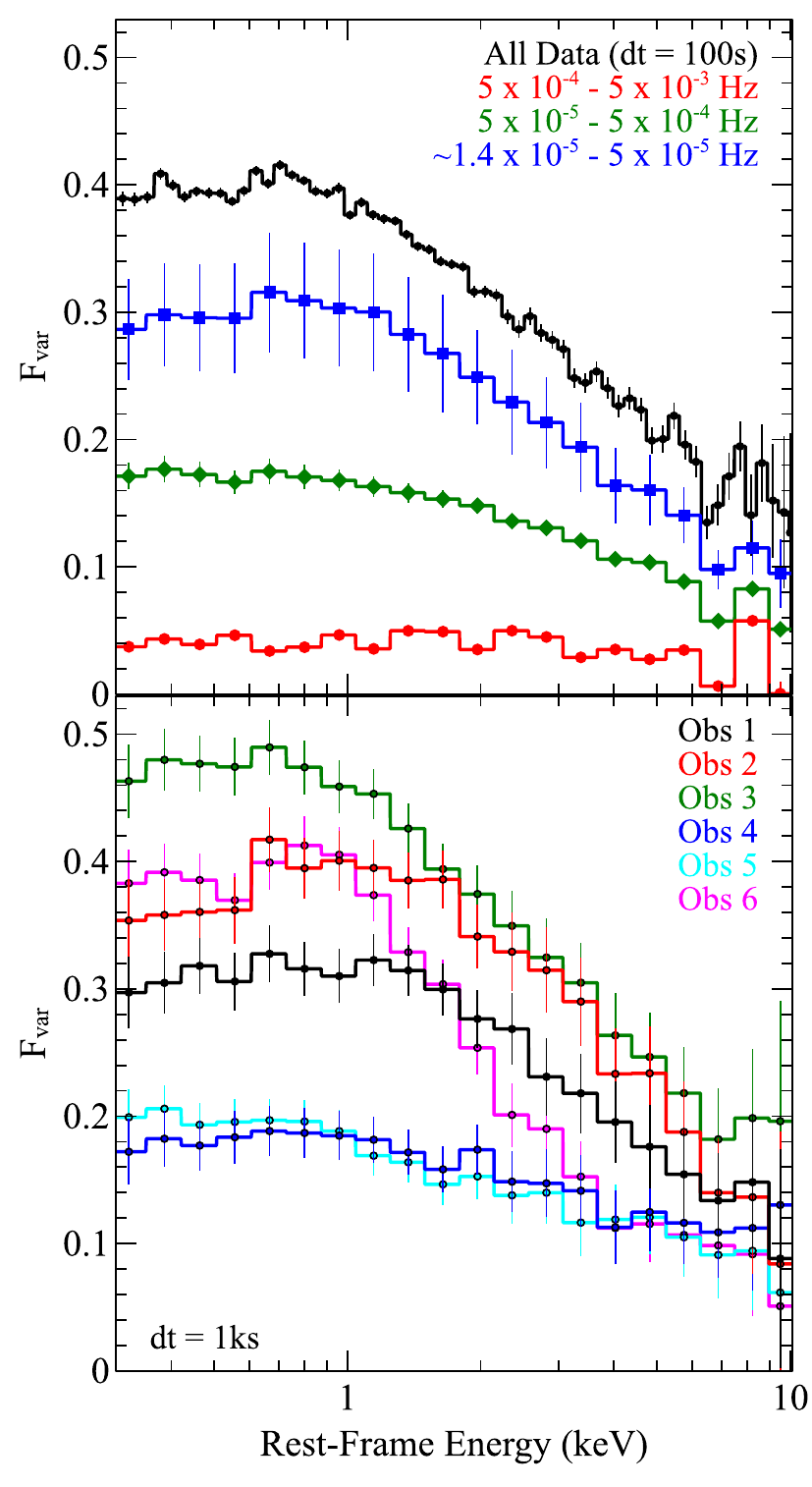}}
\end{center}
\vspace{-15pt}
\caption{The fractional rms spectra of NGC\,3227 using {\it XMM-Newton} EPIC-pn data.  Upper panel: the rms spectra are calculated in three frequency bands: $5 \times 10^{-4} - 5 \times 10^{-3}$\,Hz (red circles), $5 \times 10^{-5} - 5 \times 10^{-4}$\,Hz (green diamonds), and $\sim$1.4 $\times 10^{-5} - 5 \times 10^{-5}$\,Hz (blue squares).  The total averaged rms spectrum obtained using 100\,s time bins and 60 energy bins is shown in black.  Lower panel: the rms spectra for each of the six individual {\it XMM-Newton} observations using 1\,ks time bins.  See Section~\ref{sec:rms_spectrum} for details.}
\label{fig:rms_spectrum}
\end{figure}

We investigated the fractional rms behaviour on within-observation time-scales by computing the rms spectrum over three separate frequency bands, defined by the time-bin size, $\Delta t$, and the segment length.  Note that the upper frequency bound is given by $\nu_{\rm N} = 1/2t$ as it is set by the Nyquist frequency.  Our three frequency bands, from high-to-low, are: $5 \times 10^{-4} - 5 \times 10^{-3}$\,Hz ($\Delta t = 100$\,s; 2\,ks segments), $5 \times 10^{-5} - 5 \times 10^{-4}$\,Hz ($\Delta t = 1$\,ks; 20\,ks segments), and $\sim$1.4 $\times 10^{-5} - 5 \times 10^{-5}$\,Hz ($\Delta t = 10$\,ks; full light curves).  These are shown in red, green, and blue, respectively, in the upper panel of Fig.~\ref{fig:rms_spectrum}.  The uncertainties on each measurement are standard error on the mean obtained from averaging over the six observations.  The larger uncertainties at lower frequencies arise from the light curves have a higher variance on longer time-scales.

At the highest frequencies, the fractional rms variability is clearly low and flat across the whole bandpass.  At these frequencies, the variability is likely dominated by short-term flickering of the source and shows little energy dependence.  There is a hint of a drop in rms in the $\sim$6--7\,keV band, which may be representative of a component of Fe\,K emission that is constant on these time-scales, thus diluting the rms variability.  As we go to lower frequencies, however, the rms variability becomes significantly stronger and the spectrum steepens, with enhanced variability towards lower energies.  This suggests that the soft X-ray band largely dominates the observed variability.  This is supported by the steep photon index ($\Gamma \sim 2.1$) and component of soft excess ($kT \sim 0.1$\,keV) required by the difference spectra described in Section~\ref{sec:difference_spectra} (also see Fig.~\ref{fig:pn_nu_diff}).  Curiously, the low-frequency rms spectrum also becomes much flatter at the lowest energies; i.e. $< 1$\,keV.  Such a flat slope may be consistent with a component that varies only in flux but not spectral shape, perhaps such as a component of soft excess -- e.g. a Comptonized disc blackbody or similar -- which drives the rms spectrum at the lowest X-ray energies.  We discuss the rms behaviour of the source further in Section~\ref{sec:discussion_rms}.  We also compute the total averaged rms spectrum using all of the data across the {\it XMM-Newton} campaign with a time resolution of 100\,s.  Due to the enhanced signal by using all of the data, we can use a higher energy resolution and so we compute this rms spectrum over 60 energy bins [again, equally spaced in log$(E)$].  The errors are given by equation B2 of \citet{Vaughan03}.  Again, the shape of the rms spectrum is very steep, but also generally smooth and not particularly dominated by fine structure.  However, we again observe an apparent drop in rms in the $\sim$6--7\,keV band, likely corresponding to the Fe\,K emission complex.

We also investigated the rms behaviour on the time-scale of individual observations.  Here, we used a timing resolution of 1\,ks.  Each of the six {\it XMM-Newton} observations are shown in the lower panel of Fig.~\ref{fig:rms_spectrum}.  In general, the rms spectra are steep, again showing enhanced variability in the soft band, with a generally flattening of the slope at the lowest energies.  It is clear that some observations are more variable than others, particularly at lower energies.  For example, the steepest rms spectrum arises during obs\,3 (green) and this is most likely due to the emission flare towards the end of that observation, which is dominated by the soft band (see Fig.~\ref{fig:pn_om_lc}).  Slightly divergent behaviour can be observed in obs\,6 (magenta) as the rms spectrum appears to have a softer slope.  A different mode of variability is also evident during this observation in Fig.~\ref{fig:pn_hr_vs_flux}.  This is again consistent with the occultation event observed during this observation from a cloud of mildly ionized gas passing through the line of sight, as presented in \citet{Turner18}, which predominantly impacts the X-ray spectrum at low energies.

\subsection{The rms-flux relation} \label{sec:rms-flux}

The `rms-flux' relation shows that the rms (i.e. the absolute root-mean-square) amplitude of variability linearly scales with the X-ray flux of a source.  It is commonly observed in the X-ray variability of AGN and XRBs, but also ultraluminous X-ray sources (ULXs) and cataclysmic variables (CVs) and, in essence, shows that sources typically display stronger variability during periods when they are brighter (e.g. \citealt{UttleyMcHardy01}; \citealt{Gleissner04}; \citealt{HeilVaughan10}).

\begin{figure}
\begin{center}
\rotatebox{0}{\includegraphics[width=8.4cm]{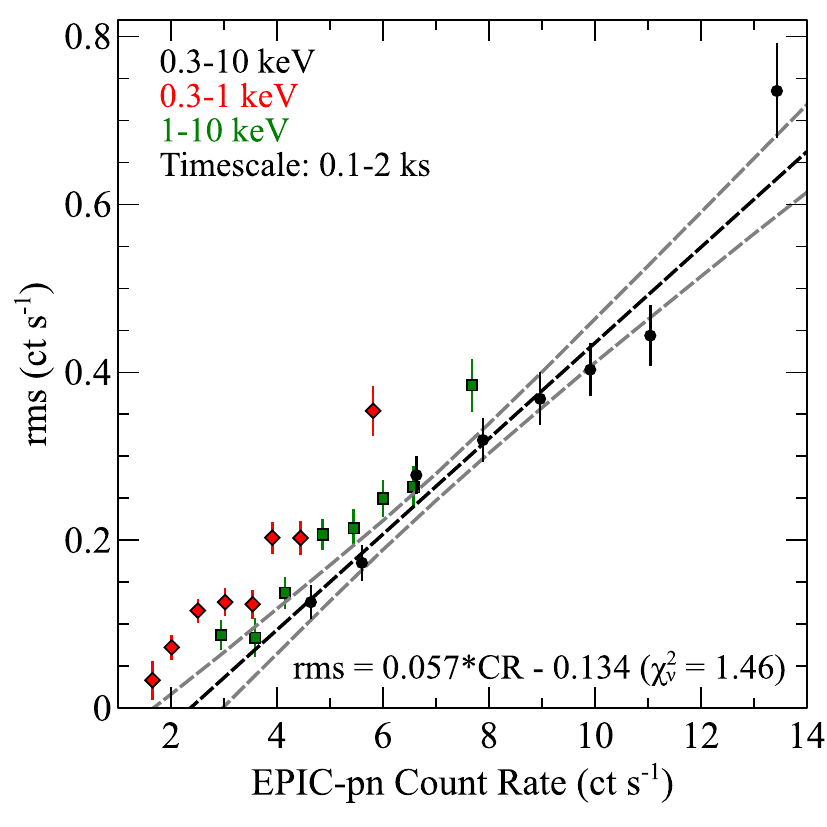}}
\end{center}
\vspace{-15pt}
\caption{The rms-flux relation of NGC\,3227 using all six {\it XMM-Newton} EPIC-pn observations.  The 0.3--10\,keV, 0.3--1\,keV, and 1--10\,keV energy bands are shown in black (circles), red (diamonds), and green (squares), respectively.  The dashed black line shows the best linear fit to the 0.3--10\,keV data, while the grey dashed lines enclose the 90\,per cent confidence band of the model.  See Section~\ref{sec:rms-flux} for details.}
\label{fig:rms-flux}
\end{figure}

To investigate this in NGC\,3227, we used all EPIC-pn data from 2016 and split the data into 2\,ks light-curve segments, using a timing resolution, $\Delta t = 50$\,s.  For each segment, we computed the both periodogram and the mean count rate.  By averaging together these periodograms over 8 count rate bins we then calculated the flux-dependent PSD.  Finally, we calculated the rms in each bin by integrating under the average PSD and subtracting the Poisson noise before taking the square root.  We calculated the rms-flux relation across three energy bands: the full 0.3--10\,keV band, a soft 0.3--1\,keV band, and hard 1--10\,keV band.\footnote{The broad-band 0.3--10\,keV rms-flux relation is lower than the sub-bands that make it up.  This may imply some degree of anticorrelation between the soft and hard bands.}  These are shown in Fig.~\ref{fig:rms-flux}.

To test for linearity in the relationship, we fitted a straight line to each of the datasets.  In the broad-band 0.3--10\,keV case, this fits the data well with a slope, $a = 0.056 \pm 0.004$ and offset, $b = -0.134 \pm 0.032$ ($\chi^{2}/{\rm d.o.f.} = 8.8/6$).  The slope corresponds to 5.6\,per cent rms on time-scales of 0.1--2\,ks.  The relationship appears to roughly hold over a factor of $> 3$ in flux.  To estimate the 90\,per cent uncertainty on this linear model, we randomly generated 1\,000 models from the parameter distribution using the best-fitting values and the covariance matrix and used the 95 and 5\,per cent rms values.  The 90\,per cent confidence band is enclosed by the grey dashed lines in Fig.~\ref{fig:rms-flux}.  

Fitting a similar linear model to the soft band returns best-fitting parameters of $a =0.064 \pm 0.006$ and $b = -0.062 \pm 0.019$ ($\chi^{2}/{\rm d.o.f.} = 11.2/6$) while, for the hard band, these values are $a = 0.059 \pm 0.006$ and $b = -0.100 \pm 0.028$ ($\chi^{2}/{\rm d.o.f.} = 5.4/6$).  So it appears that the slope is recovered and is consistent across the soft and hard bands.  We note that, in the case of each rms-flux relation, the point at which the relation intercepts the $x$-axis effectively corresponds to a component of constant emission in that given energy band.  As such, if constant components appear in both the soft and hard bands, they should co-add to reproduce the broad-band component of constant emission derived from the full-band light curve.  Here, the $x$-intercepts in the soft and hard bands correspond to $0.97 \pm 0.31$ and $1.70 \pm 0.50$\,ct\,s$^{-1}$, respectively, while in the full 0.3--10\,keV band, this value is $2.39 \pm 0.60$\,ct\,s$^{-1}$.  Combining the soft- and hard-band components yields a value of $2.67 \pm 0.59$\,ct\,s$^{-1}$, consistent with the full-band rms-flux relation.

Finally, we investigated the rms-flux behaviour within each of the individual observations and, while the relation is still positive, the relationship is difficult to constrain due to the fewer number of available segments and smaller range in flux.  Similarly, by calculating and plotting the rms against the averaged flux from each individual observation, we can also largely recover the positive relation on longer time-scales.

\subsection{The power spectrum} \label{sec:psd}

\begin{figure}
\begin{center}
\rotatebox{0}{\includegraphics[width=8.4cm]{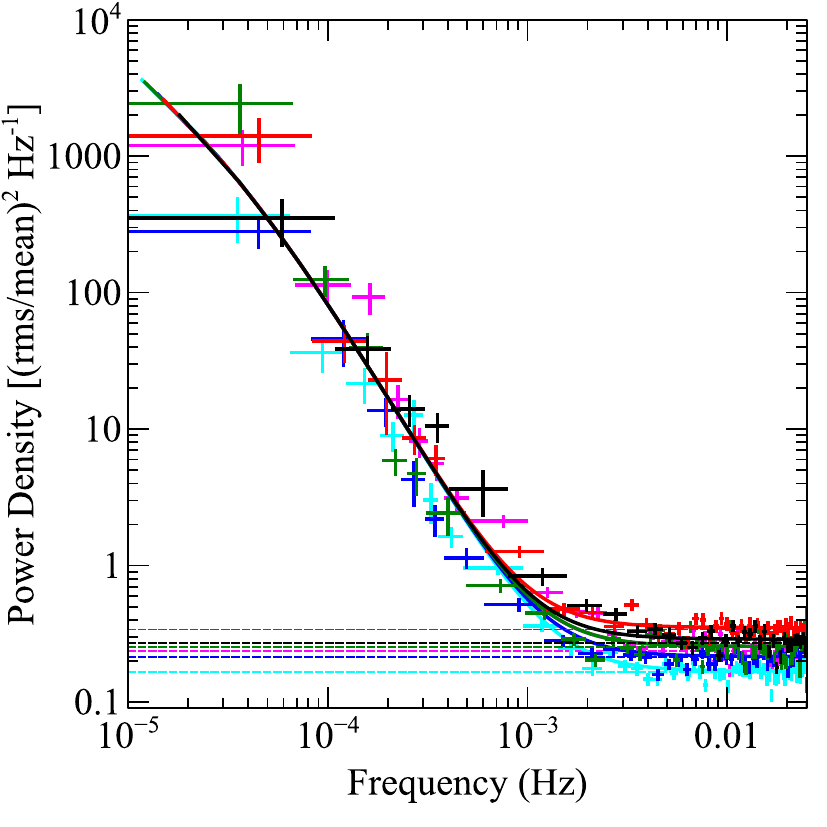}}
\end{center}
\vspace{-15pt}
\caption{The 0.3--10\,keV PSD of NGC\,3227 using all six {\it XMM-Newton} EPIC-pn observations (black, red, green, blue, cyan, magenta, respectively).  The solid line represents the best-fitting bending power-law model and the horizontal dashed lines represent the respective Poisson noise levels.  See Section~\ref{sec:psd} for details.}
\label{fig:psd}
\end{figure}

Here, we investigate the PSD of NGC\,3227.  This provides an estimate of the power of the observed variability and its dependence on temporal frequency.  EPIC-pn light curves were extracted for all six {\it XMM-Newton} observations of NGC\,3227 in 20\,s time bins.  Then, by computing simple periodograms from each orbit in units of (rms/mean)$^{2}$ (\citealt{Priestly81}; \citealt{PercivalWalden93}; \citealt{Vaughan03}), we were able to estimate the PSD down to frequencies of $\sim$10$^{-5}$\,Hz (see Fig.~\ref{fig:psd}).

The ``raw'' periodograms were fitted within \textsc{xspec}, initially with a simple model comprising a power law plus a constant:

\begin{equation} \label{eq:psd_pl} m(\nu) = N\nu^{-\alpha} + C. \end{equation}

This model has three free parameters.  These are the spectral index, $\alpha$, the power-law normalization, $N$, and an additive constant of zero slope, $C$, which accounts for the Poisson noise, which dominates at frequencies (i.e. $\gtrsim$10$^{-4}$\,Hz).  We fitted all six observations from 2016 simultaneously.  All parameters were tied together between observations except for the normalization of $C$, which we allowed to vary to account for differing levels of Poisson noise due to changes in the count rate between orbits.  To obtain the best-fitting model parameters, we minimized the Whittle statistic, $S$:

\begin{equation} \label{eq:whittle} S = 2 \sum^{N/2}_{i=1} \left\{ \frac{y_{\rm i}}{m_{\rm i}} + {\rm ln}m_{\rm i} \right\}. \end{equation}

Here, $y_{\rm i}$ is the observed value of the periodogram and $m_{\rm i}$ is the spectral density of the model at a given Fourier frequency, $\nu$ (see \citealt{Vaughan10, BarretVaughan12}).  90\,per cent confidence intervals on each parameter were estimated by finding the set of values for which $\Delta S = S(\theta) - S_{\rm min} \leq 2.7$ (where the behaviour of $\Delta S$ approximates to $\Delta \chi^{2}$).

Fitting the six observations from 0.3--10\,keV with equation~\ref{eq:psd_pl} returned best-fitting parameters of $\alpha = 2.29\pm0.08$ and log\,($N$) $= -7.3\pm0.1$.  The total number of degrees of freedom was 10\,970.  These values along with the measured Whittle statistic are provided in Table~\ref{tab:psd}.  Allowing both $N$ and $C$ to vary between datasets did improve the fit by $\Delta S  = 90$, but there was no change to the measured slope.  Additionally, allowing the fit to be driven by the low-frequency `red-noise slope' by truncating the PSDs at $10^{-4}$\,Hz returns best-fitting parameters that are consistent with those listed in Table~\ref{tab:psd}.  We also fitted equation~\ref{eq:psd_pl} to PSDs generated in `soft' and `hard'  energy bands (0.3--1 and 1--10\,keV, respectively).  One may expect to observe differences in the PSD at different energies based on the energy-dependence of the rms variability at low and high frequencies, as shown in Fig.~\ref{fig:rms_spectrum}.  For example, it is clear that larger variations are observed between the slow and fast variations in the soft band compared to the hard.  Subsequently, we may expect the soft band to have a steeper PSD.  However, we found that the PSD does not exhibit any significant energy-dependence with best-fitting values typically remaining consistent within the uncertainties.  As such, these variations are most likely contained within the uncertainties of the PSD fitting.  These values are reported in Table~\ref{tab:psd}.

\begin{table*}
\centering
\begin{tabular}{l c c c c c c c c c}
\toprule
& \multicolumn{3}{c}{Power Law} & \multicolumn{6}{c}{Bending Power Law}\\
Band (keV) & $\alpha$ & log($N$) & Whittle ($S$) & $\alpha^{{\rm low}}_{1}$ & $\nu_{b}$ ($\times 10^{-5}$\,Hz) & $\alpha^{{\rm high}}_{2}$ & log($N$) & Whittle ($S$) & $\Delta S$ \\
\midrule
Total (0.3--10\,keV) & $2.29^{+0.08}_{-0.08}$ & $-7.3^{+0.1}_{-0.1}$ & $-252\,066$ & $0.99^{+0.51}_{-0.43}$ & $3.0^{+2.4}_{-1.9}$ & $2.43^{+0.13}_{-0.12}$ & $-1.7^{+0.1}_{-0.1}$ & $-252\,076$ & 10  \\
Soft (0.3--1\,keV) & $2.27^{+0.09}_{-0.09}$ & $-7.1^{+0.1}_{-0.1}$ & $-232\,818$ & $0.98^{+0.36}_{-0.41}$ & $2.8^{+2.0}_{-1.2}$ & $2.42^{+0.14}_{-0.13}$ & $-1.8^{+0.1}_{-0.1}$ & $-232\,827$ & 9  \\
Hard (1--10\,keV) & $2.26^{+0.10}_{-0.09}$ & $-7.3^{+0.1}_{-0.1}$ & $-241\,818$ & $1.02^{+0.48}_{-0.46}$ & $3.3^{+1.8}_{-1.4}$ & $2.44^{+0.17}_{-0.15}$ & $-1.8^{+0.1}_{-0.1}$ & $-241\,828$ & 10 \\
\bottomrule
\end{tabular}
\caption{The best-fitting parameters of the two models (power-law and bending-power-law) fitted to the PSDs of NGC\,3227 in three energy bands: 0.3--10, 0.3--1 and 1--10\,keV.  The fits are applied to all six {\it XMM-Newton} EPIC-pn observations from 2016.  See Section~\ref{sec:psd} for details.}
\label{tab:psd}
\end{table*}

We then searched for evidence of a break (or bend) in the PSD by fitting a bending power law (plus a constant):

\begin{equation} P(\nu) = N\nu^{-\alpha^{low}_{1}} \left\{1+ \left\{\frac{\nu}{\nu_{b}} \right\}^{\alpha^{high}_{2}-\alpha^{low}_{1}}\right\}^{-1}, \label{eq:bendpl} \end{equation}

where $N$ is the normalization, $\nu_{b}$ is the break frequency, and $\alpha^{low}_{1}$ and $\alpha^{high}_{2}$ are the spectral indices below and above the bend, respectively (see \citealt{Gonzalez-MartinVaughan12} for further details, who find that a significant bend is detected in the PSDs of 15 out of their sample of 104 AGN observed with {\it XMM-Newton}).  Fitting equation~\ref{eq:bendpl} to the 0.3--10\,keV PSD improves the fit by $\Delta S = 10$ for two additional free parameters, suggesting the marginal detection of a bend.  The low-frequency slope, $\alpha^{low}_{1}$, is consistent with a value of $1$, consistent with the X-ray PSD slopes observed in other Seyfert galaxies.  Meanwhile, the high-frequency slope above the bend is steeper, with $\alpha^{high}_{2} = 2.43^{+0.13}_{-0.12}$, where the best-fitting bend frequency is $\nu_{b} = 3.0^{+2.4}_{-1.9} \times 10^{-5}$\,Hz, largely consistent with previous results (\citealt{UttleyMcHardy05}; \citealt{Gonzalez-MartinVaughan12}; \citealt{ArevaloMarkowitz14}).  Extending this fit to the soft and hard bands, we find consistent results ($\Delta S = 9$ and $\Delta S = 10$, respectively) with break frequencies of $\nu_{b} = 2.8^{+2.0}_{-1.2} \times 10^{-5}$ and $3.3^{+1.8}_{-1.4} \times 10^{-5}$\,Hz and high-frequency slopes of $\alpha^{high}_{2} = 2.42^{+0.14}_{-0.13}$ and $2.44^{+0.17}_{-0.15}$, respectively.  Again, the bend detection is marginal and we find no evidence of any energy-dependence of the PSD.  The best-fitting values are summarized in Table~\ref{tab:psd} and the bending-power-law fit is shown in Fig.~\ref{fig:psd}.

In Fourier-based PSD analysis, it is possible that the results may be affected by biases.  One primary form of bias is `aliasing'.  However, this has negligible effect on the data analysed here as they are contiguously sampled.  The second primary form of bias is `leakage'.  See \citet{UttleyMcHardyPapadakis02}, \citet{VaughanFabianNandra03} and \citet{Gonzalez-MartinVaughan12} - and references therein - for detailed discussion on these biases.  In terms of leakage, this may be significant when the PSD slope is intrinsically steep (e.g. $\alpha \sim 2$), potentially distorting the periodogram at the lowest observed frequencies (here: $\nu \sim 10^{-5}$\,Hz).  Subsequently, this can reduce the sensitivity to important features such as quasi-periodic oscillations and high-frequency bends while also biasing slopes which are intrinsically steep towards $\alpha \approx 2$ (see \citealt{DeeterBoynton82, UttleyMcHardyPapadakis02, VaughanFabianNandra03, Gonzalez-MartinVaughan12} for more details).

One simple method to reasonably recover accurate PSD spectral indices is `end-matching' (see \citealt{Fougere85}).  The basic end-matching process essentially consists of subtracting a linear function from the observed light curve(s).  The linear trend is defined such that the first datapoint ($y_{1}$) and the last datapoint ($y_{N}$) are joined.  Following subtraction, $y_{1} = y_{N}$.  The mean of the light curve is then restored to its original value.  We end-matched the 0.3--10\,keV EPIC-pn light curves for NGC\,3227 and re-fitted equations~(\ref{eq:psd_pl}) and~(\ref{eq:bendpl}) to the periodograms.  However, we find no significant difference to the best-fitting parameters and the results are consistent with those provided in Table~\ref{tab:psd}.

\subsection{X-ray time lags} \label{sec:lags}

Here, we analyze the X-ray Fourier time lags in NGC\,3227.  We initially use the {\it XMM-Newton} EPIC-pn data, following the methods described in the literature (i.e. \citealt{VaughanNowak97, Nowak99, Vaughan03, Uttley11, EpitropakisPapadakis16}).  Here, we can compute the cross-spectrum, allowing us to compare the variability in two separate energy bands.  The method can briefly be described as follows: (i) split the six EPIC-pn light curves into segments of identical length in two broad energy bands, (ii) compute the discrete Fourier transform for each segment, (iii) combine these, forming auto- and cross-periodograms, averaging over the number of segments.  The result is an estimate for the PSD in each band, the coherence between the two bands (see below) and phase/time lags.

\begin{figure}
\begin{center}
\rotatebox{0}{\includegraphics[width=8.4cm]{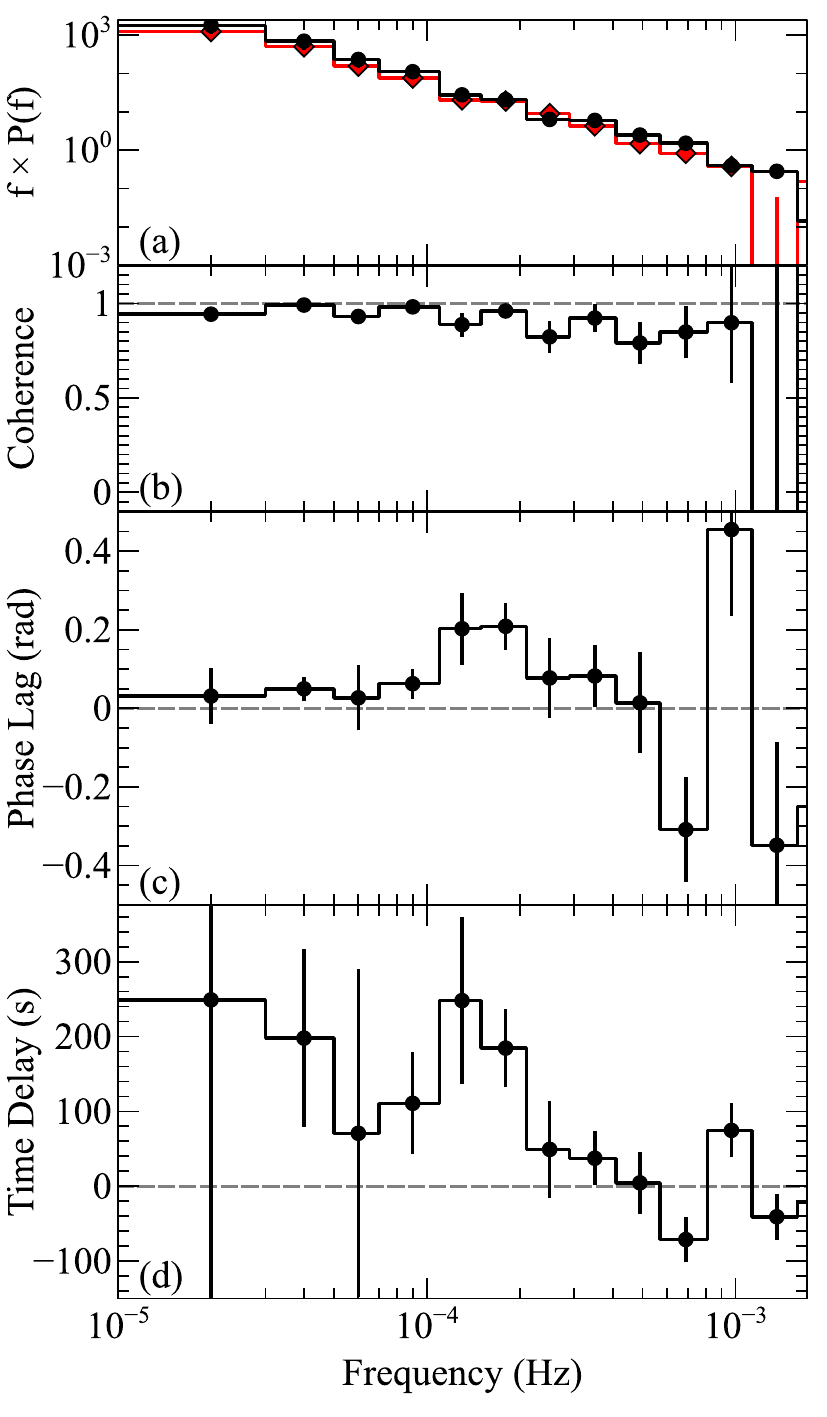}}
\end{center}
\vspace{-15pt}
\caption{The frequency-dependent timing properties of NGC\,3227 using the EPIC-pn.  Panel (a): the PSD comparing the soft (0.3--1\,keV: black) and hard (1--5\,keV: red) bands.  Panel (b): the coherence between the two energy bands.  Panel (c): the phase lag, $\phi$ (note that $\phi = \tau \times 2 \pi \nu$).  Panel (d): the time lag, $\tau$, as a function of frequency, $\nu$.  A positive phase lag denotes the hard band lagging behind the soft band.}
\label{fig:pn_time_lags}
\end{figure}

We binned up the light curve with $\Delta t = 100$\,s and used 50\,ks segment lengths, allowing us to access frequencies down to $\sim$2 $\times 10^{-5}$\,Hz.  We define our soft and hard energy bands to be 0.3--1 and 1--5\,keV, respectively.  The cross-spectral products are shown in Fig.~\ref{fig:pn_time_lags}, where the auto- and cross-periodograms are averaged over contiguous frequency bins, where each bin spans a factor of $\sim$1.4 in frequency.  Panel (a) shows the PSDs in the two energy bands, while panel (b) shows the coherence.  This is calculated from the magnitude of the periodogram \citep{VaughanNowak97}.  Its purpose is to provide a linear measurement of the correlation between the two energy bands.  The value of the coherence should fall between 0 and 1, where a value of 0 would mean that there is no correlation between the two bands, whereas a value of 1 would signify perfect coherence (i.e. the observed variations in one band would allow you to perfectly linearly predict the variations in the other band).  As such, the coherence is a valuable measurement for assessing the reality of Fourier time lags.  In the case of Fig.~\ref{fig:pn_time_lags}, it is clear that the coherence is high ($\gtrsim 0.8$) across the broad $10^{-5} - 10^{-3}$\,Hz frequency range, implying a strong correlation between the two energy bands on long time-scales (i.e. $\gtrsim 1$\,ks).  At higher frequencies ($> 10^{-3}$\,Hz), the coherence begins to rapidly drop off.  This is due to Poisson noise beginning to dominate, as is also clear from the PSD shown in Fig.~\ref{fig:psd}.

Then, in panels (c) and (d), we show the frequency-dependence of the phase lags, $\phi$, and time lags, $\tau$, respectively.  Note that the time lags are related to the phase lags by $\tau = \phi / 2\pi\nu$.  At low frequencies, a significant hard lag is observed, with the 1--5\,keV emission lagging behind the softer 0.3--1\,keV emission with time delays of up to $\sim$250\,s.  Then, as we increase in frequency, the hard lag falls off until a negative soft lag is observed at $\sim$6-8 $\times 10^{-4}$\,Hz with a measured time delay of $\tau = -70 \pm 30$\,s.

We make an attempt to assess the robustness of the soft lag measurement by following the method described in \citet{DeMarco13} and \citet{TimmerKonig95}.  Here, we employ extensive Monte Carlo simulations in order to test the reliability of the lag measurement against statistical fluctuations arising from Poisson/red noise.  Based on our fitting of the underlying PSD with a bending power law (as in Section~\ref{sec:psd}), we simulated 1\,000 pairs of stochastic light curves in the 0.3--1 and 1--5\,keV energy bands.  These were scaled to the mean count rates of the observed light curves in the same bands and were produced with the same background rates and levels of Poisson noise.  Then, for each pair of simulated light curves, we computed cross-spectral products, assuming zero phase lag (i.e. $\phi = 0$), using the same time sampling ($\Delta t = 100$\,s), segment length (50\,ks), frequency-binning (factor of 1.4) and light-curve length as we use with our real data.  As such, in our simulated cross-spectral products, we can assume that any frequency-dependent time delay arises from spurious statistical fluctuations.  Then, to test the significance of the soft lag we see in the real data, we follow the technique described in \citet{DeMarco13}.  Here, we define a `sliding-frequency' window.  This contains the same number of consecutive frequency bins, $N_{\rm w}$, as the observed lag profile.  In this instance, $N_{\rm w} = 1$.  We then compute a figure of merit [$\chi = \sum(\tau/\sigma_{\tau})^{2}$] at each step over the frequency range below those which are dominated by Poisson noise ($\lesssim 2 \times 10^{-3}$\,Hz).  In each case, we record its maximum value.  Then, the number of times that $\chi$ from the simulated data exceeds $\chi$ from our real data provides us with an estimate of the probability that such a lag could be observed by chance.  In the case of NGC\,3227, our observed soft lags are significant at a level $> 95$\,per cent.

Finally, we also explore the time lags in the harder X-ray band at energies $> 10$\,keV by utilizing the simultaneous {\it NuSTAR} data.  Due to the low-Earth orbit of the {\it NuSTAR} satellite, large gaps are introduced into the light curves.  As such, standard Fourier techniques are unsuitable.  Instead, we use a `maximum-likelihood' method, as described in \citet{Miller10a, Miller10b}.  This method rigorously accounts for `gappy' time series and allows for accurate estimates of statistical uncertainties.  Here, we create light curves in two broad energy bands and use the maximum-likelihood method to fit a joint model to the PSD in each of the bands and to the cross-spectral density.  Then, from the phases of the cross-spectral density, we can obtain time delays as a function of temporal frequency.  We note that the method was independently tested and verified by \citet{ZoghbiReynoldsCackett13}.

\begin{figure}
\begin{center}
\rotatebox{0}{\includegraphics[width=8.4cm]{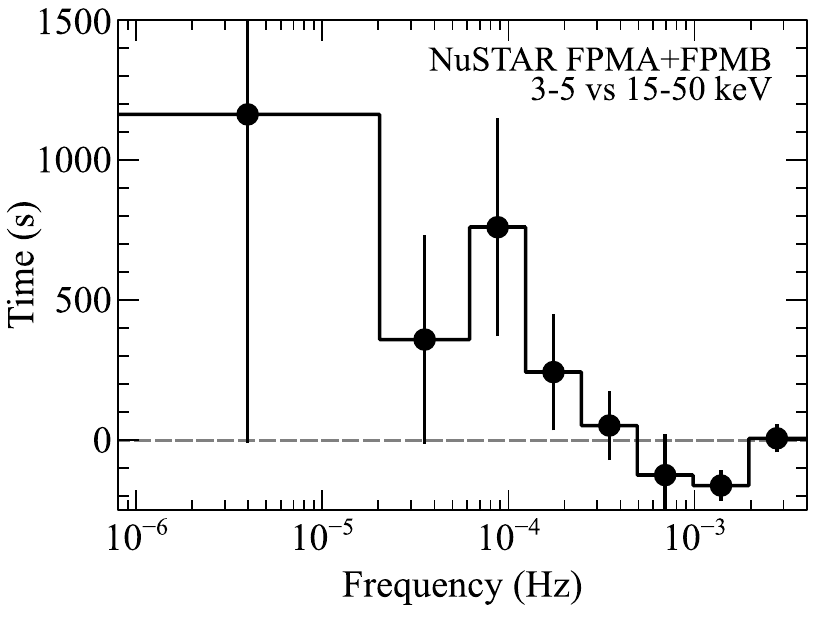}}
\end{center}
\vspace{-15pt}
\caption{The frequency-dependence of the time lags in NGC\,3227 using \textit{NuSTAR} data.  The lags are computed between the 3--5 and 15--50\,keV energy bands.  A positive lag denotes the harder band lagging behind the softer band.}
\label{fig:nustar_lags}
\end{figure}

We use all seven {\it NuSTAR} observations of NGC\,3227 and create time series with $\Delta t = 100$\,s.  Here, we focus on a softer 3--5\,keV band and a much harder 15--50\,keV band\footnote{Note that the FPM instrumental response and the shape of the source spectrum should be considered when interpreting the results.  For example, the 15--50\,keV band is very broad, but the results will be dominated by the lower end of that bandpass due to the larger statistical weight at lower energies.}, using a Fourier frequency width of $\Delta{\rm log}_{10}\nu = 0.3$.  The frequency-dependent time delays between these two bands are shown in Fig.~\ref{fig:nustar_lags}.  Again, a positive time delay indicates that the harder band is lagging behind the softer band.  Similar to the results with the {\it XMM-Newton} EPIC-pn, a hard lag is again observed at low frequencies, with the 15--50\,keV band lagging behind the 3--5\,keV band, with its magnitude rising towards lower frequencies with time delays of up to $\sim$1\,ks.  Meanwhile, at higher frequencies ($\nu \sim 5 \times 10^{-4} - 2 \times 10^{-3}$\,Hz), the lag becomes negative, with a time delay of $\sim$150\,s.  These results are largely consistent with those obtained with the EPIC-pn.  A similar result was obtained in an analysis of {\it NuSTAR} observations of NGC\,4051 \citep{Turner17}.

\subsubsection{Energy dependence of the lags} \label{sec:lag-e}

In addition to measuring the frequency-dependence of the lags, we also investigate their energy dependence, using the {\it XMM-Newton} EPIC-pn data.  Here, for a given frequency range, we compute the cross-spectral lag for a series of consecutive energy bands against a constant, broad reference band (see \citealt{Uttley11, ZoghbiUttleyFabian11, AlstonDoneVaughan14, LobbanAlstonVaughan14} for more details).  We generated cross-spectral products for nine equally-logarithmically-spaced energy bands from 0.3--10\,keV against a broad reference band, which we define to be the full 0.3--10\,keV band minus the band of interest\footnote{Note that the choice of reference band should set the (arbitrary) lag offset in the lag-energy spectrum.  To test this, we also computed lag-energy spectra using a) a constant soft reference band (0.3--1\,keV) and b) a constant harder reference band (2--5\,keV).  There were no observed changes in the shape of the resultant spectrum, but just an offset in the magnitude --- i.e. denoted by a shift on the $y$-axis.}.  So now, a positive lag denotes that the band of interest lags, on average, behind the reference band.  The errors on the lag estimates were computed using the standard procedure of \citet{BendatPiersol10}, although we note that these are expected to be conservative estimates.  This is because the light curves in each band are highly correlated and so, between adjacent energy bins, the scatter in the lags may be overestimated.  We note that a more detailed approach --- for example, if one wanted to model the lag spectrum - is described by \citet{Ingram19} to avoid over-fitting the energy-dependence of the cross-spectrum.  However, in this case, it does not have any impact on the qualitative shape of our observed lag spectrum.

\begin{figure}
\begin{center}
\rotatebox{0}{\includegraphics[width=8.4cm]{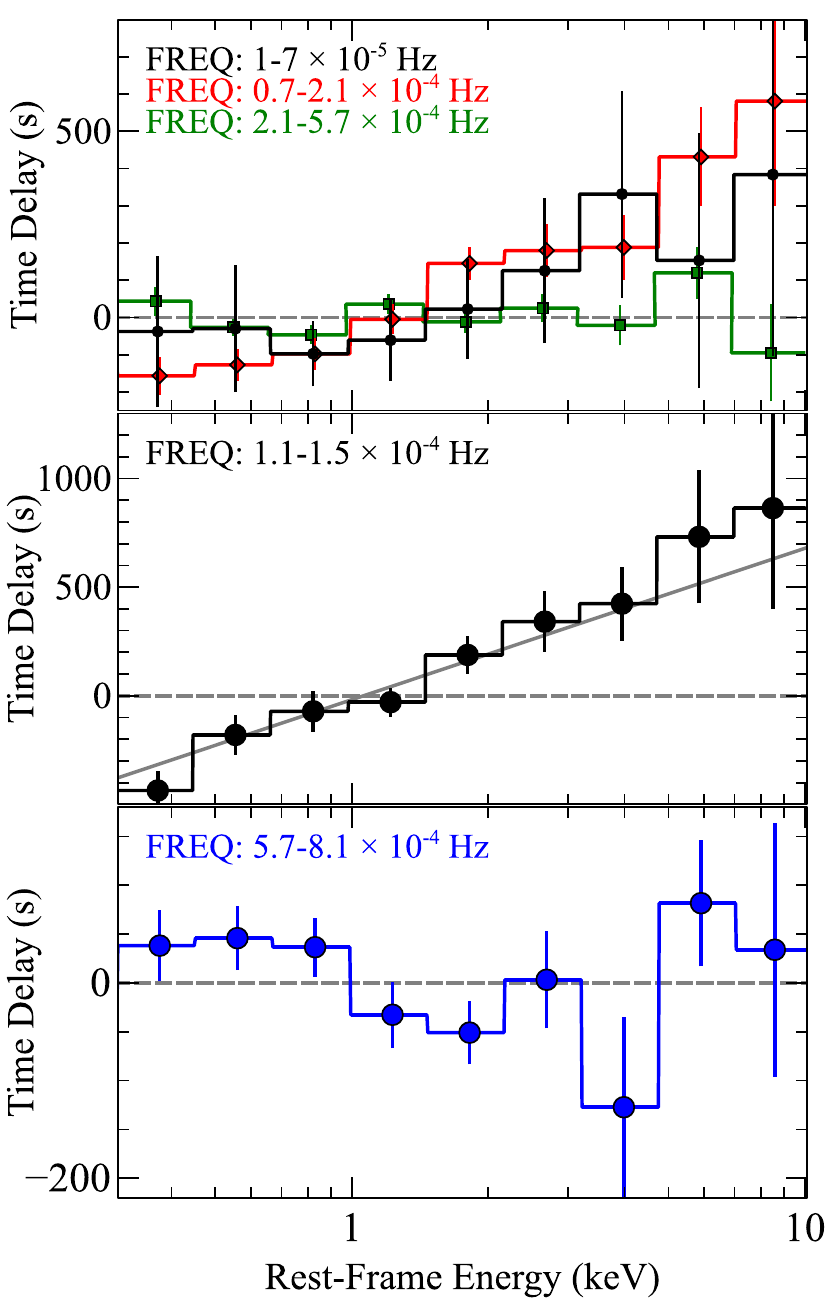}}
\end{center}
\vspace{-15pt}
\caption{The energy-dependence of the X-ray time lags in NGC\,3227 measured against a broad reference band, defined as the full 0.3--10\,keV band minus the band of interest.  Upper panel: the low-frequency hard lags in three broad frequency bands: $1-7 \times 10^{-5}$ (black circles), $0.7-2.1 \times 10^{-4}$ (red diamonds), and $2.1-5.7 \times 10^{-4}$\,Hz (green squares).  Middle and lower panels: the energy-dependence of the lags in narrower frequency bands where the hard and soft lags are most clearly defined: $1.1-1.5 \times 10^{-4}$ (black circles) and $5.7-8.1 \times 10^{-4}$\,Hz (blue circles), respectively.  The solid grey line in the middle panel shows a model fit of the form $\tau = A{\rm log}(E) + B$.}
\label{fig:pn_lag-e}
\end{figure}

The lag-energy spectra are shown in Fig.~\ref{fig:pn_lag-e}.  The upper panel plots these in three broad frequency bands $< 5.7 \times 10^{-4}$\,Hz where the hard lag emerges.  The three frequency bands are: $1-7 \times 10^{-5}$, $0.7-2.1 \times 10^{-4}$ and $2.1-5.7 \times 10^{-4}$\,Hz.  These correspond to splitting the nine lowest frequency bins in Fig.~\ref{fig:pn_time_lags} into three groups of three.  It is clear that strong energy-dependence of the lags is observed in the two lowest frequency bands, with the magnitude of the hard lag increasing towards higher energies, peaking at $\tau \sim 500$\,s with respect to the broad reference band.  At the highest of these three frequency bins, the energy-dependence drops off as the lag approaches zero.  In the middle panel, we focus in on a narrower frequency band, where the hard lag is most clearly observed.  This is the $1.1-1.5 \times 10^{-4}$\,Hz frequency range.  Here, the energy-dependence is very clearly defined, with a maximum delay in the hardest band (i.e. the 7--10\,keV band lags behind the 0.3--7\,keV band at $\nu = 1.1-1.5 \times 10^{-4}$\,Hz with a time delay of $\tau = 864 \pm 465$\,s).  We note that the shape of the lag-energy spectrum is suggestive of a log-linear dependence, whereby $\tau$ scales roughly linearly with log($E$).  Therefore, we fitted the lag-energy spectrum with a model taking the form: $\tau = A{\rm log}(E) + B$, where $E$ is the energy band and $A$ and $B$ are constants.  Here, we find that $A = 303 \pm 45$ and $B = -18 \pm 34$, with $\chi^{2}/{\rm d.o.f.} = 10.0/8$.  The model fit is overlaid in Fig.~\ref{fig:pn_lag-e}.  Such a log-linear dependence is seen in various AGN (e.g. Ark\,564: \citealt{Kara13}; IRAS\,18325$-$5926: \citealt{LobbanAlstonVaughan14}; and PG\,1211$+$143: \citealt{Lobban18a}) and XRBs (e.g. Cygnus\,X-1: \citealt{Nowak99} and GX\,339-4: \citealt{Uttley11}) and is often discussed in terms of the propagating fluctuations model (see \citealt{KotovChurazovGilfanov01}).  However, we note that such a model does not account for existence of high-frequency soft lags.

The lowest panel of Fig.~\ref{fig:pn_lag-e} focuses on the $5.7-8.1 \times 10^{-4}$\,Hz frequency range, where the high-frequency soft lag emerges [see Fig.~\ref{fig:pn_time_lags}: panel (d)].  Here, the energy-dependence is not so clearly defined.  From $\lesssim 4$\,keV, there is a tentative hint of the lag increasing in magnitude towards lower energies, as seen in other sources (e.g. \citealt{Kara13, Lobban18a}), although its significance is marginal here.  We note that the magnitude of the high-frequency lag does increase at energies $> 5$\,keV.  While this tentatively appears similar in shape to the Fe\,K lags reported in various other sources (e.g. \citealt{AlstonDoneVaughan14, Kara14}), such a feature does not appear to be significantly detected here.  This is perhaps consistent with the lack of any significantly variable component of Fe\,K$\alpha$ emission in the X-ray spectrum.

\subsubsection{Visualizing the low-frequency time delays in smoothed light curves}

\begin{figure}
\begin{center}
\rotatebox{0}{\includegraphics[width=8.4cm]{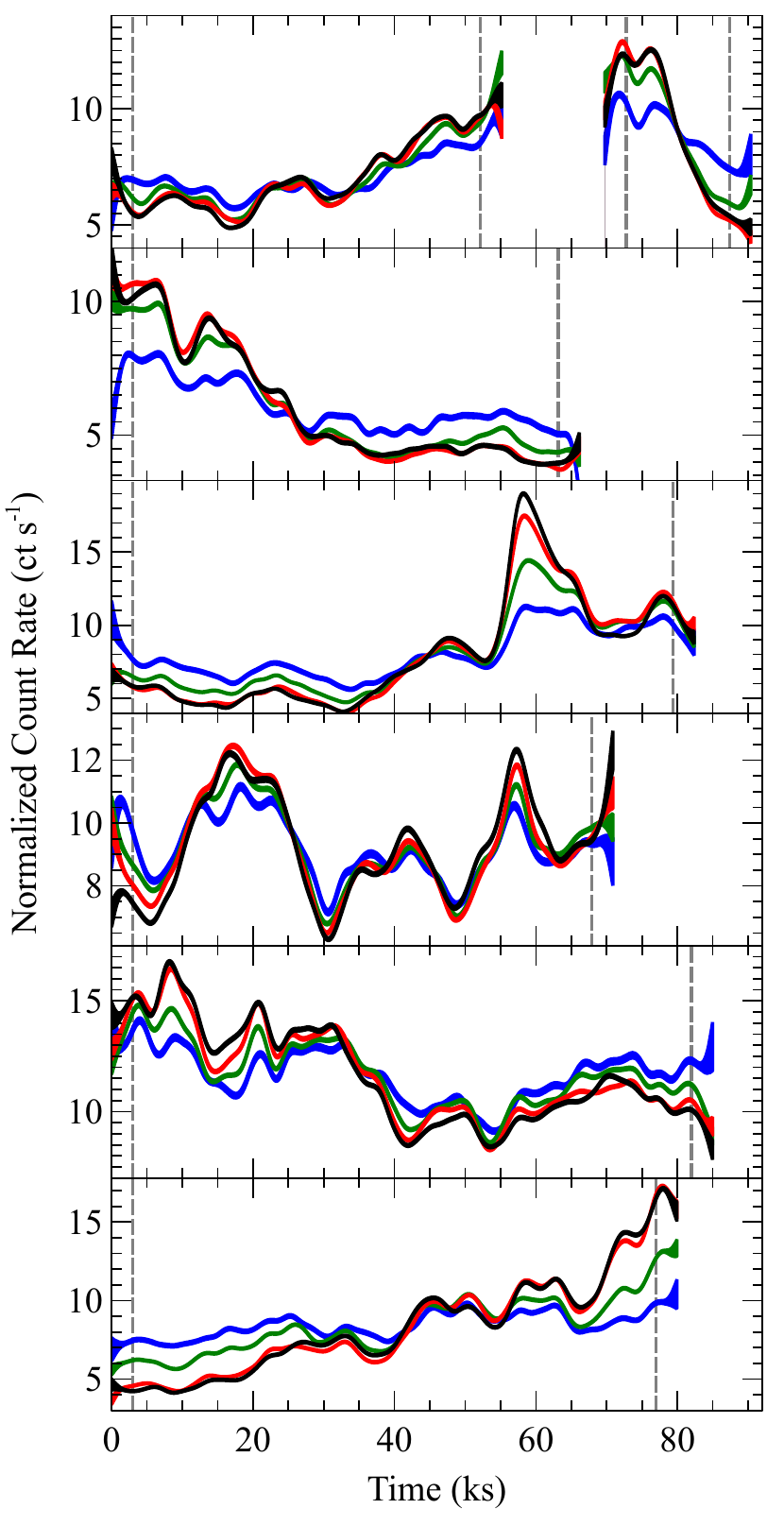}}
\end{center}
\vspace{-15pt}
\caption{The EPIC-pn light curves of NGC\,3227, smoothed via convolution with a Gaussian ($\sigma = 2$\,ks).  The curves are normalized to their mean 0.3--10\,keV count rates and plotted in four energy bands: 0.3--0.7 (black), 0.7--1.5 (red), 1.5--5 (green) and 5--10\,keV (blue).  Meanwhile, the thickness of the bands denote the 90\,per cent confidence intervals.  The vertical dashed lines mark the edges of the convolution kernel (i.e. $3 \times \sigma = 6$\,ks) and, so, beyond these limits, we assume the light curves begin to become unreliable.}
\label{fig:smoothed_lc}
\end{figure}

In addition to measuring time lags in the Fourier domain, it may be also useful to look for evidence of them --- and other aspects of low-frequency energy-dependent behaviour --- in the time domain (e.g. see \citealt{Lobban18a}, where we discuss a similar approach in the context of PG\,1211$+$143, finding curious changing-lag behaviour).  One possibility is to smooth out the high-frequency variability such that the low-frequency variations are all that remain.  In Fig.~\ref{fig:smoothed_lc}, we show the results of such an approach, whereby we smoothed the 50\,ks-binned EPIC-pn light curves by convolving them with a Gaussian of width $\sigma = 2$\,ks.  We did try other Gaussian widths, but found that $\sigma = 2$\,ks gave the clearest results, given the rapid variability of the source.  We smoothed light curves in four energy bands: 0.3--0.7, 0.7--1.5, 1.5--5 and 5--10\,keV.  To better compare the curves, we then normalized each light curve to the mean 0.3--10\,keV EPIC-pn count rate for each given observation.  Finally, we computed 90\,per cent confidence bands for each light curve by performing 10\,000 simulations based on the observed count rate in each bin and plotting the 5 and 95\,per cent boundaries from the resultant count-rate distribution.

In the case of NGC\,3227, the variations and subsequent time delays are rapid (i.e. $\tau \sim$ few hundred seconds) and so the low-frequency lags are not obviously apparent in the light curves.\footnote{There are instances in which a low-frequency hard lag is potentially visible --- e.g. in obs\,3 where the soft-band minima (black and red) can be seen to lead the hard-band minima (green and blue) at $\sim$16 and $\sim$33\,ks.  A clear hard lag is also visible in the hardest band during the minima at $\sim$42\,ks in obs\,5.}  However, other curious behaviour is apparent.  In particular, the emission flare during obs\,3 ($\sim$60\,ks) shows strong energy-dependent behaviour.  It is clear that the flare is dominated by enhanced emission in the softest X-ray bands, while the hardest band (5--10\,keV) only increases in flux slightly before reaching a plateau.  Similar, but more moderate-strength energy-dependence can be seen in obs\,5 during the flare occurring at the beginning of the observation.  Meanwhile, in obs\,6, it is clear that the soft-band emission (0.3--1.5\,keV inclusive) is significantly more variable than the hard band, with its steady increase in flux with respect to its harder counterpart apparent in the hardness ratio analysis presented in Figs~\ref{fig:pn_om_lc} and~\ref{fig:pn_hr_vs_flux}.  See \citet{Turner18} for a detailed spectral analysis of this observation.

\section{Discussion}

We have presented a series of fundamental variability properties of the highly-variable AGN, NGC\,3227, through a long {\it XMM-Newton} $+$ {\it NuSTAR} observing campaign.  Below, we discuss the results.

\subsection{Spectral decomposition}

In Sections~\ref{sec:lightcurves},~\ref{sec:spectral_decomposition}, and~\ref{sec:difference_spectra}, we explored the spectral variations of NGC\,3227.  The energy-dependent spectral evolution of the source was tracked through hardness ratio measurements, where we find that the source displays typical softer-when-brighter behaviour.  This is particularly evident during the large flare midway through obs\,3, which is dominated by an increase in soft-band emission.  Curiously, we find evidence for the source to be typically harder than average during obs\,6, while the X-ray spectrum gradually becomes softer over the course of $\lesssim 1$\,day.  This appears to be due to a rapid occultation event of the central continuum source arising from the passage of a mildly-ionized cloud of gas ($N_{\rm H} \sim 10^{22}$\,cm$^{-2}$) across the line of sight.  This is discussed in detail in \citet{Turner18}.

In Fig.~\ref{fig:high_mid_low_flux_spectra}, we fit three flux-selected spectra in order to explore the spectral variability.  We apply the baseline model from the spectral analysis by \citet{Turner18} and find that the bulk of the variability appears to be dominated by changes in the strength of the primary power-law continuum with mild changes in photon index (ranging from $\Gamma = 1.4-1.7$), providing further evidence of softer-when-brighter source behaviour.  Superimposed on this are weak variations of $\sim$20\,per cent in the strength of the neutral reflection component, with the bulk of the modest variability in the {\it NuSTAR} bandpass simply accounted for by changes in the power-law continuum.  Meanwhile, we find that the component of Fe\,K$\alpha$ emission at $\sim$6.4\,keV shows hints of flux-variability on these time-scales consistent with the weak variability of the neutral reflection component at the 2$\sigma$ level.  This modest variability, relative to the strong continuum variability, is suggestive of an origin in material that is distant from the central X-ray source.

Finally, in Fig.~\ref{fig:pn_nu_diff}, we show the broad-band difference spectra of NGC\,3227.  While the hard X-rays remain largely invariant in spectral shape, just with modest differences in normalization, it is clear that large deviations from an absorbed power-law fit are apparent at low energies ($\lesssim 2$\,keV), indicative of additional modes of variability.  These residuals are accounted for by a multiplicative component of warm absorption and a soft excess, emerging at energies $< 1$\,keV.  The difference spectra are dominated by a steep power-law-like component ($\Gamma \sim 2.1$), further indicative of enhanced variability in the soft band.  This is also borne out by the rms variability analysis (discussed below).  The difference spectra show very little excess emission in the Fe\,K band or at energies $> 10$\,keV.  This supports a picture whereby the neutral reflector and associated Fe\,K$\alpha$ emission originate in distant material.  Given the time-scale of this observing campaign, this would correspond to a distance of $> 1$\,light-month from the central black hole.

\subsection{The rms variability} \label{sec:discussion_rms}

In Section~\ref{sec:rms_spectrum}, we show the rms variability of NGC\,3227 and its associated energy dependence.  On short time-scales ($0.1-2$\,ks: $\nu = 5 \times 10^{-4} - 5 \times 10^{-3}$\,Hz), the magnitude of the variability is found to be low, with a flat dependence on energy --- i.e. the low-energy and high-energy X-rays all typically vary by $\sim$5\,per cent rms on time-scales $< 2$\,ks.  As such, on these time-scales, the emission in the hard and soft bands varies with roughly the same fractional amplitude\footnote{Conversely, for example, if a hard or soft component varied, this would manifest itself as a drop in the rms at low or high energies, respectively, where dilution from the component of constant flux becomes significant.}.

As we move towards longer time-scales ($2-20$\,ks), however, enhanced variability begins to emerge in the soft band with its magnitude increasing towards lower energies.  This suggests that the primary source of spectral variability in this observing campaign from 2016 may be due to a soft power-law-like component, varying in flux, that is steeper than the $\Gamma \sim 1.7$ power law that dominates the hard band.  We recently observed similar behaviour in the sources PG\,1211$+$143 \citep{Lobban16} and Ark\,120 \citep{Lobban18b}.  In this scenario, the X-ray continuum can largely be described as a blend of two components where the soft excess varies slowly and independently of the harder X-ray coronal power law, similar to the suggestion by \citet{ArevaloMarkowitz14} from a previous 100\,ks {\it XMM-Newton} observation of NGC\,3227 (also see Ton\,S180: \citealt{Edelson02}; Ark\,564: \citealt{Turner01}; and Mkn\,509: \citealt{Mehdipour11}).  If Comptonization is the mechanism responsible for producing the variable soft-excess component, one possibility is that this arises from intrinsic coronal variations, either in terms of the electron temperature or optical depth.  At the lowest frequencies (20-$\sim$70\,ks), the rms variability is stronger and its spectral shape becomes steeper.  Curiously, at low energies ($\sim$0.3--1.0\,keV), the rms spectrum is observed to flatten out, suggestive of additional complexity in the long-time-scale variability.  A flat shape could arise from a soft component which varies in flux but whose spectral shape remains largely invariant.  Such a component could not extend to higher energies here, though, or the rms would be equally as high, unless it was diluted by a constant hard component which dominates at higher energies.  However, as the hard reflection component is weak in this source (just $\sim$10\,per cent of the overall flux from 0.3--50\,keV), this scenario can be ruled out.  Instead, it may be the case that, at low frequencies, the variability from 1--10\,keV is dominated by a steep power-law-like component (i.e. whose intrinsic variability increases with decreasing energy), while the rms variability $< 1$\,keV may be dominated by a component of soft excess varying in flux while maintaining a steady spectral shape.  We note that, superimposed on these broad-band spectral changes, the source exhibits additional variability due to line-of-sight variations in the warm absorber.  However, these absorption changes do not appear to significantly contribute to the averaged rms spectrum, which may be the case if the warm absorber variations occur on longer time-scales than probed here.  We also compute the overall time-averaged rms spectrum using all of the data with a timing resolution of 100\,s time bins and finer energy resolution.  Again, the rms spectrum is steep, while it becomes flatter at energies $< 1$\,keV.  On the whole, the rms spectrum is smooth with limited discrete structure.  However, we do observe a small drop in the fractional rms at $\sim$6--7\,keV, which coincides with the emission line from near-neutral Fe\,K$\alpha$.  This apparent drop in rms could potentially arise due to dilution from a quasi-constant component of Fe\,K emission --- i.e. if the bulk of the emission originates far from the black hole.  Indeed, in the time-averaged spectrum, the Fe\,K$\alpha$ emission line contributes roughly 10\,per cent of the total observed flux from 6--7\,keV.  Given the variability and flux level of NGC\,3227, if such a component were constant on these time-scales, it would likely produce a drop of $\sim$10--15\,per cent in the fractional rms spectrum, roughly consistent with what we observe.  Finally, in the lower panel of Fig.~\ref{fig:rms_spectrum}, we show the rms spectra from the six individual {\it XMM-Newton} observations of NGC\,3227 using a time resolution of 1\,ks.  All six rms spectra are steep, although it is clear that some observations are more variable than others.  In particular, the emission flare during obs\,3 results in a steeper rms spectrum. Meanwhile, divergent behaviour can be observed during obs\,6, which coincides with the absorption event described in \citet{Turner18} in which a cloud of mildly ionized gas is observed the occult our line of sight to the central source.

In Section~\ref{sec:rms-flux}, we also explore the rms variability of NGC\,3227 in terms of its flux-dependence.  It is clear from Fig.~\ref{fig:rms-flux} that the source displays enhanced variability during periods of brighter flux.  The observed rms-flux relation is roughly linear, which is consistent with the rapid behaviour variability observed in other accreting sources across a wide range of luminosities and masses, such as AGN, XRBs, ultraluminous X-ray sources and cataclysmic variables (e.g. \citealt{UttleyMcHardy01, Gleissner04, HeilVaughan10, Scaringi12}).  A linear rms-flux relation is often discussed in terms of propagating fluctuations, where the X-ray emission and mass-accretion-rate variations are multiplicatively coupled \citep{UttleyMcHardyVaughan05} - although this is not necessarily a unique explanation of this relation.  We note that a similarly linear relationship is observed in soft (0.3--1\,keV) and hard (1--10\,keV) bands (albeit covering a smaller range in flux) and so a linear rms-flux relationship appears to hold across the EPIC-pn X-ray bandpass.

\subsection{The power spectrum}

In Section~\ref{sec:psd}, we computed and modelled the broad-band 0.3--10\,keV PSD of NGC\,3227.  We find a marginal detection of a bend in the PSD ($\Delta S = 10$ for two additional free parameters) with a best-fitting bend frequency of $\nu_{\rm b} = 3.0^{+2.4}_{-1.9} \times 10^{-5}$\,Hz.  The low-frequency slope, $\alpha^{low}_{1}$, is consistent with a value of 1, while the high-frequency slope above the bend is much steeper: $\alpha^{high}_{2} = 2.43^{+0.13}_{-0.12}$.  We find consistent results fitting the PSD in the soft (0.3--1\,keV) and hard (1--10\,keV) bands and, hence, no energy-dependence of the PSD.

To date, a number of discrepant values have been reported in the literature for the bend frequency in NGC\,3227.  \citet{Gonzalez-MartinVaughan12} modelled the PSD from a single $\sim$100\,ks {\it XMM-Newton} observation, finding a best-fitting bend frequency of $\nu_{\rm b} = 2.3 \pm 0.7 \times 10^{-4}$\,Hz from 0.2--10\,keV, which is a factor of $\sim$8 higher than our result.  However, \citet{UttleyMcHardy05} and \citet{KellySobolewskaSiemiginowska11} independently analysed the PSD using {\it XMM-Newton} and {\it RXTE} data from 2--10\,keV, reporting bend frequencies of $\sim$2.6 $\times 10^{-5}$ and $\sim$3.7 $\times 10^{-5}$\,Hz, respectively.  These results are consistent with those we report here.

A relatively linear relationship between the bend time-scale of the PSD and the mass of the black hole might be expected based on simple scaling arguments \citep{Fender07}. Indeed, such a correlation has been observed for a number of AGN (e.g. \citealt{UttleyMcHardyPapadakis02, Markowitz03}).  In the case of NGC\,3227, our observed bend frequency corresponds to a time-scale of $T_{\rm b} = 1/\nu_{\rm b} = 33^{+21}_{-26}$\,ks (or $0.39^{+0.24}_{-0.31}$\,days).  A simple scaling relation between the time-scale of the bend and the mass of the black hole is provided by \citet{Gonzalez-MartinVaughan12}:

\begin{equation} {\rm log}(T_{\rm b}) = A{\rm log}(M_{\rm BH}) + C. \label{eq:psd_bend} \end{equation}

Here, $T_{\rm b}$ is the time-scale of the break measured in days and $M_{\rm BH}$ is the mass of the black hole in units of $\times 10^{6}$\,$M_{\odot}$, while and $A$ and $C$ are coefficients.  In the case of NGC\,3227, we assume $M_{\rm BH} = 6.0^{+1.2}_{-1.4} \times 10^{6}$\,$M_{\odot}$ \citep{BentzKatz15}.  Meanwhile, \citet{Gonzalez-MartinVaughan12} derive values of $A = 1.09 \pm 0.21$ and $C = -1.70 \pm 0.29$ from fitting equation~\ref{eq:psd_bend} to a sample of Seyfert galaxies.  This predicts the time-scale of the bend to lie in the range: $T_{\rm b} = 0.03 - 0.25$\,days. Our observed value of $T_{\rm b} = 0.39^{+0.24}_{-0.31}$\,days is consistent with the existing scaling relation.

Additionally, \citet{McHardy06} provide an extension to the mass-time-scale relation by including a dependence on the bolometric luminosity, $L_{\rm bol}$ (also see \citealt{Kording07}). In this instance, equation~\ref{eq:psd_bend} is modified to include an an additional term: $B{\rm log}(L_{\rm bol})$.  Here, $L_{\rm bol}$ is in units of $10^{44}$\,erg\,s$^{-1}$, while the best-fitting values of the coefficients are derived by \citet{Gonzalez-MartinVaughan12} to be $A = 1.34 \pm 0.36$, $B = -0.24 \pm 0.28$ and $C = -1.88 \pm 0.36$. For NGC\,3227, this predicts a value of $T_{\rm b} = 0.16$\,days (although with large uncertainties), where we assume $L_{\rm bol} = 7 \times 10^{43}$\,erg\,s$^{-1}$ (\citealt{WooUrry02}).  This is also consistent with what we observe within our measurement uncertainties.

\subsection{X-ray time lags}

In Section~\ref{sec:lags}, we investigated the short-time-scale variability of NGC\,3227 through X-ray Fourier time lags.  Through analysis of {\it XMM-Newton} EPIC-pn data, we find a low-frequency hard lag, whereby the 1--5\,keV band emission lags behind the softer 0.3--1\,keV emission with time delays of up to a few hundred seconds.  The magnitude of the lag appears to increase towards lower frequencies.  Additionally, we observe a roughly log-linear energy-dependence of the hard lag that is also borne out in our analysis of simultaneous {\it NuSTAR} data, where we show that the lag extends into the higher-energy bandpass, with the 15--50\,keV emission lagging behind the 3--5\,keV emission with time delays of up to $\sim$1\,ks (see Section~\ref{sec:lag-e}).  Low-frequency hard lags may be ubiquitous in accreting black-hole systems with the amplitude and frequency of the lags scaling with $M_{\rm BH}$, ranging from XRBs to AGN.  As such, they are likely an important phenomenon, carrying crucial information regarding the structure of black-hole accretion flows.  Following our approach in \citet{Lobban18a}, we do make an attempt to visualize the low-frequency lags in the time domain through creating smoothed light curves (see Fig.~\ref{fig:smoothed_lc}), although the small time delays (i.e. $\tau \sim$ few hundred s) are difficult to pick out.  Nevertheless, we observe enhanced emission in the soft band during a strong flare in obs\,3 and we also demonstrate the gradual softening of the spectrum during obs\,6 as the spectrum uncovers over the course of $\lesssim 1$\,day following a rapid occultation event \citep{Turner18}.

We also detect a significant high-frequency soft lag in NGC\,3227.  This manifests itself in the EPIC-pn data in the $\sim$6--8 $\times 10^{-4}$\,Hz frequency range with the 0.3--1\,keV emission delayed with respect to the 1--5\,keV emission with a time delay of $\tau = -70 \pm 30$\,s.  A scaling relation was reported by \citet{DeMarco13} linking the amplitude/frequency of the soft lag to the mass of the black hole.  This was based on a sample of 15 AGN in which high-frequency soft lags have been detected. They report the following relations between the observed frequency, $\nu$, time lag, $\tau$, and the black hole mass: ${\rm log}(\nu) = -3.50[\pm 0.07] - 0.47[\pm 0.09] {\rm log}(M_{\rm BH})$ and ${\rm log}(\mid\tau\mid) = 1.98[\pm 0.08] + 0.59[\pm 0.11] {\rm log}(M_{\rm BH})$, where $M_{\rm BH}$ is mass of the black hole in units of $10^{7}$\,$M_{\odot}$. For the mass estimate of NGC\,3227 ($6^{+1.4}_{-1.2} \times 10^{6}$\,$M_{\odot}$), these relations predict the soft-lag frequency to lie in the range: $3.1- 4.5 \times 10^{-4}$\,Hz, while the time delay falls in the range: $36-58$\,s.  Therefore, while our observed soft lag does occur at a slightly higher frequency than predicted by existing scaling relations, the measured time delay is consistent with the expected value.

A number of models exist that attempt to explain the origin of the lags, ranging from propagating fluctuations \citep{Lyubarskii97} to small-scale reverberation (e.g. \citealt{ZoghbiUttleyFabian11, Fabian13}) to secondary Comptonization components (e.g. as per \citealt{Done12, GardnerDone14}), which may be associated with the outer layers of the accretion disc.  For any model, it is important to explain the full spectrum of lags observed across a broad range of frequencies.

A popular model to explain the origins of low-frequency hard lags in XRBs is the `propagating fluctuations' model \citep{Lyubarskii97}.  Due to the behavioural similarities exhibited by the hard lags in XRBs (e.g. Cygnus X-1: \citealt{KotovChurazovGilfanov01}) and numerous variable AGN (e.g. \citealt{McHardy04, Fabian13, LobbanAlstonVaughan14}), it has been argued that a comparable mechanism is responsible in all accreting black hole systems.  However, this model does not account for the existence of the high-frequency soft lags and so, explaining the time delays over a broad frequency range then requires a two-component model additionally involving small-scale reverberation of the primary X-rays by matter close to the SMBH, perhaps due to reflection (e.g. \citealt{ZoghbiUttleyFabian11, Fabian13}).  However, we note that there does not appear to be any requirement for any strong component of ionized reflection (blurred or otherwise) in NGC\,3227 (e.g. \citealt{Markowitz09, Beuchert15, Turner18}), although there is likely a requirement for a neutral reflector producing a modest Compton reflection hump $> 10$\,keV.  We also do not find any strong evidence of any high-frequency Fe\,K lags, often strongly associated with the small-scale reverberation model.

In addition, by using `maximum-likelihood' methods to measure the time lags in the simultaneous {\it NuSTAR} data, we also find that the soft lag exists (in the same frequency band as in the EPIC-pn data) between the 3--5 and 15--50\,keV bands.  Given that the softer reference band (3--5\,keV) contains a much lower fraction of reflected emission compared to the hard 15--50\,keV band (where the Compton reflection hump becomes apparent), it is difficult to envisage a mechanism whereby the softer 3--5\,keV {\it NuSTAR} band could lag behind the harder band via small-scale reverberation, similar to the case of NGC\,4051 (see \citealt{Turner17}).

In \citet{Miller10a, Miller10b}, a reverberation model is presented in which the lags emerge via scattering of the primary X-rays by circumnuclear material.  This photoelectrically-absorbing material is typically placed at tens to hundreds of $r_{\rm g}$ from the central black hole and may be associated with a disc wind/outflow (also see \citealt{Mizumoto18, Mizumoto19}).  Here, the low-frequency hard lags are produced by scattering from this absorbing material while the higher-frequency soft lags may arise from oscillatory, ringing features, which are a consequence of taking the Fourier transform of a sharp feature in the time domain (see \citealt{Turner17} for a detailed discussion).  In this scenario, the fraction of scattered-to-direct light increases with energy, predicting a strong delayed signal in the hard X-ray band.  \citet{Turner17} analysed a series of {\it NuSTAR} observations of the highly-variable Seyfert galaxy NGC\,4051, discovering that a soft lag exists between the softer 2--4, 5--7.5 and 8--15\,keV bands and the hard 15--70\,keV band.  Here, the transition from the low-frequency hard lags to the higher-frequency soft lag is accounted for by reverberation from circumnuclear material with the high scattered fraction of light indicating a global covering fraction of the reprocessor of $\sim$50\,per cent with respect to the continuum source.  As shown in Fig.~\ref{fig:pn_lag-e}, we also observe the time delays increasing towards higher energies. in the case of NGC\,3227, ultimately peaking in the {\it NuSTAR} bandpass.  Evidence for such circumnuclear material in NGC\,3227 is well established through previous X-ray studies (e.g. \citealt{LamerUttleyMcHardy03, Markowitz09, Markowitz14, RiversMarkowitzRothschild11}) with the source occasionally observed to undergo line-of-sight variable absorption events (e.g. \citealt{LamerUttleyMcHardy03, Beuchert15}).  Such an occultation event was observed in this very campaign, as shown in \citet{Turner18} and the spectral decomposition presented in Section~\ref{sec:spectral_decomposition}.  As such, it is also conceivable that we are seeing evidence of these `light echoes' in this source from absorbing material tens to hundreds of $r_{\rm g}$ from the primary continuum source.

\subsection{Summary} \label{sec:summary}

In this paper, we have presented a series of X-ray variability results from a long {\it XMM-Newton} and {\it NuSTAR} observing campaign on the bright AGN NGC\,3227.  We find the source to exhibit strong variability in both the X-ray and UV bands.  The typical trend is for the source to be steeper when brighter, consistent with other similar-class AGN.  However, NGC\,3227 does also undergo a period of significant spectral hardening due to an occultation event by a cloud of mildly ionized gas passing into the line of sight.  This largely manifests itself as absorption signatures --- primarily from the Fe UTA - in the soft X-ray band.

We spectrally decompose the source and show that the primary components that comprise the broad-band X-ray spectrum are a power law ($\Gamma \sim 1.7$), a modest neutral reflection component with an associated narrow Fe\,K$\alpha$ emission line, an additional component of soft excess and additional zones of absorption arising from mildly outflowing ($v_{\rm out} \sim$ 100-1\,000\,km\,s$^{-1}$) ionized gas.  The warm absorber is observed to exhibit moderate variability, primarily in terms of changes in the line-of-sight column density.  Meanwhile, the bulk of the observed variability appears to be driven by the continuum, whose magnitude scales with the flux of the source in a roughly linear way.  The broad-band X-ray variability shows strong energy dependence.  On time-scales of 0.1--2\,ks, the variability is weak and displays roughly the same fractional amplitude across the bandpass.  On longer time-scales of 2--20\,ks, enhanced variability is observed in the soft band, consistent with the spectral variations being dominated by a steep spectral component.  This trend continues to longer time-scales of 20--$\sim$70\,ks, although with a curious flattening of the fractional variability amplitude at the lowest energies.  One possible cause of this is that the soft band ($\lesssim 1$\,keV) variability becomes dominated by a component of soft excess --- e.g. a Comptonized disc blackbody --- which varies in flux while remaining roughly invariant in spectral shape.  We note that the variability of the reflection component, however, is weak ($\sim$20\,per cent) on all time-scales, suggesting an origin in material that is distant from the central black hole. 

Finally, we employ Fourier methods and find a marginal detection of a bend in the PSD.  Given the black hole mass of NGC\,3227, the slopes of the low- and high-frequency parts of the PSD and the bend time-scale are consistent with existing scaling relations in the literature.  The variability of the soft and hard bands are well correlated in this source over a large range of frequencies, with a high coherence.  As such, we compute X-ray time lags, finding a hard lag at low frequencies and a soft lag with a time delay of $\tau = -70 \pm 30$\,s at higher frequencies ($\nu \sim 6-8 \times 10^{-4}$\,Hz), roughly consistent with the predictions from existing scaling relations.  Through maximum-likelihood methods, we extend our analysis to the {\it NuSTAR} bandpass, finding that the hard 15--50\,keV band lags behind the softer 3--5\,keV band.  Given that the reflection component in NGC\,3227 is weak (with no requirement for any component of relativistically-blurred ionized reflection), and that the softer reference band is much more continuum-dominated, it is difficult to reconcile this as originating via small-scale reverberation.  Instead, given the well-established evidence for the existence of large quantities of circumnuclear photoelectrically-absorbing material a few tens to hundreds of $r_{\rm g}$ from the black hole in NGC\,3227, we instead consider that these lags may arise from this material via energy-dependent scattering.  Such material may typically be associated with a disc wind or outflow.

\section*{Acknowledgements}

This research has made use of the NASA Astronomical Data System (ADS), the NASA Extragalactic Database (NED), and is based on observations obtained with (i) the {\it XMM-Newton} satellite, an ESA science mission with instruments and contributions directly funded by ESA Member States and the USA (NASA) and (ii) the {\it NuSTAR} mission, a project led by the California Institute of Technology (Caltech) and managed by the Jet Propulsion Laboratory (JPL).  AL is a current ESA research fellow and also acknowledges support from the UK STFC under grant no. ST/M001040/1.  TJT acknowledges NASA grant no. NNX17AD91G. VB acknowledges financial support through NASA grant no. NNX17AC40G and through the CSST Visiting Scientist Initiative.  We also thank our anonymous referee for a careful and thorough review of this paper.


{}


\begin{thebibliography}{}

\bibitem[\protect\citeauthoryear{Alston, Done \& Vaughan}{Alston et al.}{2014}]{AlstonDoneVaughan14}Alston W. N., Done C., Vaughan S., 2014, MNRAS, 439, 1548

\bibitem[\protect\citeauthoryear{Ar\'{e}valo et al.}{2008}]{Arevalo08}Ar\'{e}valo P., Uttley P., Kaspi S., Breedt E., Lira P., McHardy I. M., 2008, MNRAS, 389, 1479

\bibitem[\protect\citeauthoryear{Ar\'{e}valo \& Markowitz}{2014}]{ArevaloMarkowitz14}Ar\'{e}valo P., Markowitz A., 2014, ApJ, 783, 83

\bibitem[\protect\citeauthoryear{Arnaud}{1996}]{Arnaud96}Arnaud K. A., 1996, in Jacoby G. H., Barnes J., eds, ASP Conf. Ser. Vol. 101, Astronomical Data Analysis Software and Systems V. Astron. Soc. Pac., San Francisco, p. 17

\bibitem[\protect\citeauthoryear{Barret \& Vaughan}{2012}]{BarretVaughan12}Barret D., Vaughan S., 2012, ApJ, 746, 131

\bibitem[\protect\citeauthoryear{Bendat \& Piersol}{2010}]{BendatPiersol10}Bendat J. S., Piersol A. G., 2010, Random Data: Analysis and Measurement Procedures, 4th edn. Wiley, New York

\bibitem[\protect\citeauthoryear{Bentz \& Katz}{2015}]{BentzKatz15}Bentz M. C., Katz S., 2015, PASP, 127, 67

\bibitem[\protect\citeauthoryear{Beuchert et al.}{2015}]{Beuchert15}Beuchert T. et al., 2015, A\&A, 584, A82

\bibitem[\protect\citeauthoryear{Cui et al.}{1997}]{Cui97}Cui W., Zhang, S. N., Focke, W., Swank, J. H., 1997, ApJ, 484, 383

\bibitem[\protect\citeauthoryear{Deeter \& Boynton}{1982}]{DeeterBoynton82}Deeter J. E. \& Boynton P. E., 1982, ApJ, 261, 337

\bibitem[\protect\citeauthoryear{De Marco et al.}{2013}]{DeMarco13}De Marco B., Ponti G., Cappi M., Dadina M., Uttley P., Cackett E. M., Fabian A. C., Miniutti G., 2013, MNRAS, 431, 2441

\bibitem[\protect\citeauthoryear{de Vaucouleurs et al.}{1991}]{deVaucouleurs91}de Vaucouleurs G., de Vaucouleurs A., Corwin H. G., Jr., Buta R. J., Paturel G., Fouqu\'{e}, P. 1991, Third Reference Catalogue of Bright Galaxies (New York: Springer)

\bibitem[\protect\citeauthoryear{Done et al.}{2012}]{Done12}Done C., Davis S. W., Jin C., Blaes O., Ward M., 2012, MNRAS, 420, 1848

\bibitem[\protect\citeauthoryear{Edelson et al.}{2002}]{Edelson02}Edelson R. et al., 2002, ApJ, 568, 610

\bibitem[\protect\citeauthoryear{Epitropakis \& Papadakis}{2016}]{EpitropakisPapadakis16}Epitropakis A., Papadakis I. E., 2016, A\&A, 591, 113

\bibitem[\protect\citeauthoryear{Fabian et al.}{2013}]{Fabian13}Fabian A. C. et al., 2013, MNRAS, 429, 2917

\bibitem[\protect\citeauthoryear{Fender et al.}{2007}]{Fender07}Fender R., Koerding E., Belloni T., Uttley P., McHardy I., Tzioumis T., 2007, preprint (arXiv:0706.3838)

\bibitem[\protect\citeauthoryear{Fougere}{1985}]{Fougere85}Fougere P. F., 1985, JGR, 90, 4355

\bibitem[\protect\citeauthoryear{Gardner \& Done}{2014}]{GardnerDone14}Gardner E., Done C., 2014, MNRAS, 442, 2456

\bibitem[\protect\citeauthoryear{Gaskell}{2004}]{Gaskell04}Gaskell M., 2004, ApJL, 612, L21

\bibitem[\protect\citeauthoryear{Gleissner et al.}{2004}]{Gleissner04}Gleissner T. et al., 2004, A\&A, 425, 1061

\bibitem[\protect\citeauthoryear{Gonz{\'a}lez-Mart{\'i}n \& Vaughan}{2012}]{Gonzalez-MartinVaughan12}Gonz{\'a}lez-Mart{\'i}n O., Vaughan S., 2012, A\&A, 544, 80

\bibitem[\protect\citeauthoryear{Haardt \& Maraschi}{1993}]{HaardtMaraschi93}Haardt F., Maraschi I., 1993, ApJ, 413, 507

\bibitem[\protect\citeauthoryear{Harrison et al.}{2013}]{Harrison13}Harrison F. A. et al., 2013, ApJ, 770, 103

\bibitem[\protect\citeauthoryear{Heil \& Vaughan}{2010}]{HeilVaughan10}Heil L. M., Vaughan S., 2010, MNRAS, 405, 86

\bibitem[\protect\citeauthoryear{Ingram}{2019}]{Ingram19}Ingram A., 2019, MNRAS, 489, 3927

\bibitem[\protect\citeauthoryear{Jansen et al.}{2001}]{Jansen01}Jansen F. et al., 2001, A\&A, 365, 1

\bibitem[\protect\citeauthoryear{Jin, Done \& Ward}{Jin et al.}{2017}]{JinDoneWard17}Jin C., Done C., Ward M., 2017, MNRAS, 468, 3663

\bibitem[\protect\citeauthoryear{Kalberla et al.}{2005}]{Kalberla05}Kalberla P. M. W., Burton W. B., Hartmann D., Arnal E. M., Bajaja E., Morras R., P\"{o}ppel W. G. L., 2005, A\&A, 440, 775

\bibitem[\protect\citeauthoryear{Kara et al.}{2013}]{Kara13}Kara E., Fabian A. C., Cackett E. M., Uttley P., Wilkins D. R., Zoghbi A., 2013, MNRAS, 434, 1129

\bibitem[\protect\citeauthoryear{Kara et al.}{2014}]{Kara14}Kara E., Cackett E. M., Fabian A. C., Reynolds C., Uttley P., 2014, MNRAS, 439, L26

\bibitem[\protect\citeauthoryear{Kara et al.}{2016}]{Kara16}Kara E., Alston W. N., Fabian A. C., Cackett E. M., Uttley P., Reynolds C. S., Zoghbi A., 2016, MNRAS, 462, 511

\bibitem[\protect\citeauthoryear{Kallman \& Bautista}{2001}]{KallmanBautista01}Kallman T., Bautista M., 2001, ApJS, 133, 221

\bibitem[\protect\citeauthoryear{Kallman et al.}{2004}]{Kallman04}Kallman T. R., Palmeri P., Bautista M. A., Mendoza C., Krolik J. H., 2004, ApJS, 155, 675

\bibitem[\protect\citeauthoryear{Kelly, Sobolewska \& Siemiginowska}{Kelly et al.}{2011}]{KellySobolewskaSiemiginowska11}Kelly, B. C., Sobolewska, M., Siemiginowska, A., 2011, ApJ, 730, 52

\bibitem[\protect\citeauthoryear{K\"{o}rding et al.}{2007}]{Kording07}K\"{o}rding E. G., Migliari S., Fender R., Belloni T., Knigge C., McHardy I., 2007, MNRAS, 380, 301

\bibitem[\protect\citeauthoryear{Kotov, Churazov \& Gilfanov}{Kotov et al.}{2001}]{KotovChurazovGilfanov01}Kotov O., Churazov E., Gilfanov M., 2001, MNRAS, 327, 799

\bibitem[\protect\citeauthoryear{Krongold et al.}{2007}]{Krongold07}Krongold Y., Nicastro F., Elvis M., Brickhouse N., Binette L., Mathur S., Jim\'{e}nez-Bail\'{o}n E., 2007, ApJ, 659, 1022

\bibitem[\protect\citeauthoryear{Lamer, Uttley \& McHardy}{Lamer et al.}{2003}]{LamerUttleyMcHardy03}Lamer G., Uttley P., McHardy I. M., 2003, MNRAS, 342, L41

\bibitem[\protect\citeauthoryear{Lawrence et al.}{1987}]{Lawrence87}Lawrence A., Watson M. G., Pounds K. A., Elvis M., 1987, Nature, 325, 694

\bibitem[\protect\citeauthoryear{Lobban, Alston \& Vaughan}{Lobban et al.}{2014}]{LobbanAlstonVaughan14}Lobban A. P., Alston W. N., Vaughan S., 2014, MNRAS, 445, 3229

\bibitem[\protect\citeauthoryear{Lobban et al.}{2016}]{Lobban16}Lobban A. P., Vaughan S., Pounds K., Reeves J. N., 2016, MNRAS, 457, 38

\bibitem[\protect\citeauthoryear{Lobban et al.}{2018a}]{Lobban18a}Lobban A. P., Vaughan S., Pounds K., Reeves J. N., 2018a, MNRAS, 476, 225

\bibitem[\protect\citeauthoryear{Lobban et al.}{2018b}]{Lobban18b}Lobban A. P., Porquet D., Reeves J. N., Markowitz A., Nardini E., Grosso N., 2018b, MNRAS, 474, 3237

\bibitem[\protect\citeauthoryear{Lyubarskii}{1997}]{Lyubarskii97}Lyubarskii Y. E., 1997, MNRAS, 292, 679

\bibitem[\protect\citeauthoryear{Magdziarz \& Zdziarski}{1995}]{MagdziarzZdziarski95}Magdziarz P., Zdziarski A. A., 1995, MNRAS, 273, 837

\bibitem[\protect\citeauthoryear{Markowitz et al.}{2003}]{Markowitz03}Markowitz A. et al., 2003, ApJ, 593, 96

\bibitem[\protect\citeauthoryear{Markowitz et al.}{2009}]{Markowitz09}Markowitz A., Reeves J. N., George I. M., Braito V., Smith R., Vaughan S., Ar\'{e}valo P., Tombesi F., 2009, ApJ, 691, 922

\bibitem[\protect\citeauthoryear{Markowitz et al.}{2014}]{Markowitz14}Markowitz A., Krumpe M., Nikutta R., 2014, MNRAS, 439, 1403

\bibitem[\protect\citeauthoryear{Mason et al.}{2001}]{Mason01}Mason K. O. et al., 2001, A\&A, 365, 3

\bibitem[\protect\citeauthoryear{McHardy et al.}{2004}]{McHardy04}McHardy I. M., Papadakis I. E., Uttley P., Page M. J., Mason K. O., 2004, MNRAS, 348, 783

\bibitem[\protect\citeauthoryear{McHardy et al.}{2006}]{McHardy06}McHardy I. M., Koerding E., Knigge C., Uttley P., Fender R. P., 2006, Nature, 444, 730

\bibitem[\protect\citeauthoryear{McHardy et al.}{2007}]{McHardy07}McHardy I. M., Ar\'{e}valo P., Uttley P., Papadakis I. E., Summons D. P., Brinkmann W., Page M. J., 2007, MNRAS, 382, 985

\bibitem[\protect\citeauthoryear{Mehdipour et al.}{2011}]{Mehdipour11}Mehdipour M. et al., 2011, A\&A, 534, 39

\bibitem[\protect\citeauthoryear{Miller et al.}{2010a}]{Miller10a}Miller L., Turner T. J., Reeves J. N., Lobban A., Kraemer S. B., Crenshaw D. M., 2010a, MNRAS, 403, 196

\bibitem[\protect\citeauthoryear{Miller et al.}{2010b}]{Miller10b}Miller L., Turner T. J., Reeves J. N., Braito V., 2010b, MNRAS, 408, 1928

\bibitem[\protect\citeauthoryear{Miyamoto \& Kitamoto}{1989}]{MiyamotoKitamoto89}Miyamoto S., Kitamoto S., 1989, Nature, 342, 773

\bibitem[\protect\citeauthoryear{Mizumoto et al.}{2018}]{Mizumoto18}Mizumoto M., Done C., Hagino K., Ebisawa K., Tsujimoto M., Odaka H., 2018, MNRAS, 478, 971

\bibitem[\protect\citeauthoryear{Mizumoto et al.}{2019}]{Mizumoto19}Mizumoto M., Ebisawa K., Tsujimoto M., Done C., Hagino K., Odaka H., 2019, MNRAS, 2019, 482, 5316

\bibitem[\protect\citeauthoryear{Mould et al.}{2000}]{Mould00}Mould J.R. et al., 2000, ApJ, 529, 786

\bibitem[\protect\citeauthoryear{Nandra et al.}{1997}]{Nandra97}Nandra K., George I. M., Mushotzky R. F., Turner T. J., Yaqoob T., 1997, ApJ, 476, 70

\bibitem[\protect\citeauthoryear{Nandra et al.}{2007}]{Nandra07}Nandra K., O'Neill P. M., George I. M., Reeves J. N., 2007, MNRAS, 382, 194

\bibitem[\protect\citeauthoryear{Nowak et al.}{1999}]{Nowak99}Nowak M. A., Vaughan B. A., Wilms J., Dove J. B., Begelman M. C., 1999, ApJ, 510, 874

\bibitem[\protect\citeauthoryear{Papadakis}{2004}]{Papadakis04}Papadakis I. E., 2004, MNRAS, 348, 207

\bibitem[\protect\citeauthoryear{Papadakis, Nandra \& Kazanas}{Papadakis et al.}{2001}]{PapadakisNandraKazanas01}Papadakis I. E., Nandra K., Kazanas D., 2001, ApJ, 554, 133

\bibitem[\protect\citeauthoryear{Park et al.}{2006}]{Park06}Park T., Kashyap V. L., Siemiginowska A., van Dyk D. A., Zezas A., Heinke C., Wargelin B. J., 2006, ApJ, 652, 610

\bibitem[\protect\citeauthoryear{Percival \& Walden}{1993}]{PercivalWalden93}Percival D. B., \& Walden A. T., 1993, Spectral analysis for physical applications: multi taper and conventional univariate techniques, Cambridge University Press, Cambridge

\bibitem[\protect\citeauthoryear{Priestly}{1981}]{Priestly81}Priestly M. B., 1981, Spectral Analysis and Time Series, Academic Press, London

\bibitem[\protect\citeauthoryear{Rivers, Markowitz \& Rothschild}{Rivers et al.}{2011}]{RiversMarkowitzRothschild11}Rivers E., Markowitz A., Rothschild R., 2011, ApJS, 193, 3

\bibitem[\protect\citeauthoryear{Scaringi et al.}{2012}]{Scaringi12}Scaringi S., Koerding E., Uttley P., Knigge C., Groot P. J., Still M., 2012, MNRAS, 421, 2854

\bibitem[\protect\citeauthoryear{Shakura \& Sunyaev}{1973}]{ShakuraSunyaev73}Shakura N. I., Sunyaev R. A., 1973, A\&A, 24, 337

\bibitem[\protect\citeauthoryear{Shang et al.}{2011}]{Shang11}Shang Z. et al., 2011, ApJS, 196, 2

\bibitem[\protect\citeauthoryear{Str\"{u}der et al.}{2001}]{Struder01}Str\"{u}der L. et al., 2001, A\&A, 365, L18

\bibitem[\protect\citeauthoryear{Timmer \& K\"{o}nig}{1995}]{TimmerKonig95}Timmer J., K\"{o}nig M., 1995, A\&A, 300, 707

\bibitem[\protect\citeauthoryear{Turner et al.}{2001}]{Turner01}Turner T. J., Romano P., George I. M., Edelson R., Collier S. J., Mathur S., Peterson B. M., 2001, ApJ, 561, 131

\bibitem[\protect\citeauthoryear{Turner et al.}{2017}]{Turner17}Turner T. J., Miller L., Reeves J. N., Braito V., 2017, MNRAS, 467, 3924

\bibitem[\protect\citeauthoryear{Turner et al.}{2018}]{Turner18}Turner T. J., Reeves J. N., Braito V., Lobban A., Kraemer S., Miller L., 2018, MNRAS, 481, 2470

\bibitem[\protect\citeauthoryear{Uttley \& McHardy}{2001}]{UttleyMcHardy01}Uttley P., McHardy I. M., MNRAS, 2001, 323, 26

\bibitem[\protect\citeauthoryear{Uttley \& McHardy}{2005}]{UttleyMcHardy05}Uttley P., McHardy I. M., MNRAS, 363, 586

\bibitem[\protect\citeauthoryear{Uttley, McHardy \& Vaughan}{Uttley et al.}{2005}]{UttleyMcHardyVaughan05}Uttley P., McHardy I. M., Vaughan S., 2005, MNRAS, 359, 345

\bibitem[\protect\citeauthoryear{Uttley, McHardy \& Papadakis}{2002}]{UttleyMcHardyPapadakis02}Uttley P., McHardy I. M., Papadakis I. E., 2002, MNRAS, 332, 231

\bibitem[\protect\citeauthoryear{Uttley et al.}{2011}]{Uttley11}Uttley P., Wilkinson T., Cassatella P., Wilms J., Pottschmidt K., Hanke M., B\"{o}ck M., 2011, MNRAS, 414, 60

\bibitem[\protect\citeauthoryear{Uttley et al.}{2014}]{Uttley14}Uttley P., Cackett E. M., Fabian A. C., Kara E., Wilkins D. R., 2014, A\&ARv, 22, 72

\bibitem[\protect\citeauthoryear{Vaughan \& Nowak}{1997}]{VaughanNowak97}Vaughan B. A., Nowak M. A., 1997, ApJ, 474, 43

\bibitem[\protect\citeauthoryear{Vaughan et al.}{2003}]{Vaughan03}Vaughan S., Edelson R., Warwick R. S., Uttley P., 2003, MNRAS, 345, 1271

\bibitem[\protect\citeauthoryear{Vaughan, Fabian \& Nandra}{Vaughan et al.}{2003}]{VaughanFabianNandra03}Vaughan S., Fabian A. C., Nandra K., 2003, MNRAS, 339, 1237

\bibitem[\protect\citeauthoryear{Vaughan}{2010}]{Vaughan10}Vaughan S., 2010, MNRAS, 402, 307

\bibitem[\protect\citeauthoryear{Verner et al.}{1996}]{Verner96}Verner D. A., Ferland G. J., Korista K. T., Yakovlev D. G., 1996, ApJ, 465, 487

\bibitem[\protect\citeauthoryear{Willingale et al.}{2013}]{Willingale13}Willingale R., Starling R. L. C., Beardmore A. P., Tanvir N. R., O'Brien P. T., 2013, MNRAS, 431, 394

\bibitem[\protect\citeauthoryear{Wilms, Allen \& McCray}{Wilms et al.}{2000}]{WilmsAllenMcCray00}Wilms J., Allen A., McCray R., 2000, ApJ, 542, 914

\bibitem[\protect\citeauthoryear{Woo \& Urry}{2002}]{WooUrry02}Woo J-H., Urry M. C., 2002, ApJ, 579, 530

\bibitem[\protect\citeauthoryear{Zoghbi, Uttley \& Fabian}{Zoghbi et al.}{2011}]{ZoghbiUttleyFabian11}Zoghbi A., Uttley P., Fabian A. C., 2011, MNRAS, 412, 59

\bibitem[\protect\citeauthoryear{Zoghbi, Reynolds \& Cackett}{Zoghbi et al.}{2013}]{ZoghbiReynoldsCackett13}Zoghbi A., Reynolds C., Cackett E. M., 2013, ApJ, 777, 24

\end{thebibliography}
\end{document}